\documentclass[12pt]{spieman}  % 12pt font required by SPIE;
\usepackage{amsmath,amsfonts,amssymb}
\usepackage{graphicx}
\usepackage{setspace}
\usepackage{tocloft}
\usepackage[breakwords]{truncate}

\usepackage{multicol}
\usepackage{subcaption}
\usepackage[acronym, toc]{glossaries}
\usepackage[font={small}]{caption}
\usepackage{fancyhdr}

\glsdisablehyper
\let\olditemize\itemize
\def\itemize{\olditemize\itemsep=-4pt }
\let\oldenumerate\enumerate
\def\enumerate{\oldenumerate\itemsep=-3pt }
\usepackage[right]{lineno}

\title{KM3NeT front-end and readout electronics system: hardware, firmware and software}
\author[a]{S.~Aiello}
\author[b]{F.~Ameli}
\author[c]{M.~Andre}
\author[d]{G.~Androulakis}
\author[e]{M.~Anghinolfi}
\author[f]{G.~Anton}
\author[g]{M.~Ardid}
\author[h]{J.~Aublin}
\author[d]{C.~Bagatelas}
\author[i,j]{G.~Barbarino}
% ...

\affil[a]{INFN, Sezione di Catania, Via Santa Sofia 64, Catania, 95123 Italy}
\affil[b]{INFN, Sezione di Roma, Piazzale Aldo Moro 2, Roma, 00185 Italy}
\affil[c]{Universitat Polit{\`e}cnica de Catalunya, Laboratori d'Aplicacions Bioac{\'u}stiques, Centre Tecnol{\`o}gic de Vilanova i la Geltr{\'u}, Avda. Rambla Exposici{\'o}, s/n, Vilanova i la Geltr{\'u}, 08800 Spain}
\affil[d]{NCSR Demokritos, Institute of Nuclear and Particle Physics, Ag. Paraskevi Attikis, Athens, 15310 Greece}
\affil[e]{INFN, Sezione di Genova, Via Dodecaneso 33, Genova, 16146 Italy}
\affil[f]{Friedrich-Alexander-Universit{\"a}t Erlangen-N{\"u}rnberg, Erlangen Centre for Astroparticle Physics, Erwin-Rommel-Stra{\ss}e 1, 91058 Erlangen, Germany}
\affil[g]{Universitat Polit{\`e}cnica de Val{\`e}ncia, Instituto de Investigaci{\'o}n para la Gesti{\'o}n Integrada de las Zonas Costeras, C/ Paranimf, 1, Gandia, 46730 Spain}
\affil[h]{APC, Universit{\'e} Paris Diderot, CNRS/IN2P3, CEA/IRFU, Observatoire de Paris, Sorbonne Paris Cit\'e, 75205 Paris, France}
\affil[i]{INFN, Sezione di Napoli, Complesso Universitario di Monte S. Angelo, Via Cintia ed. G, Napoli, 80126 Italy}
\affil[j]{Universit{\`a} di Napoli ``Federico II'', Dip. Scienze Fisiche ``E. Pancini'', Complesso Universitario di Monte S. Angelo, Via Cintia ed. G, Napoli, 80126 Italy}
% ...

\makeglossaries

\newacronym{CPU}{CPU}{Central Processing Unit}
\newacronym{FPGA}{FPGA}{Field Programmable Gate Array}
\newacronym{LVDS}{LVDS}{Low Voltage Differential Signaling}
\newacronym{DOM}{DOM}{Digital Optical Module}
\newacronym{DAC}{DAC}{Digital to Analog Converter}
\newacronym{DU}{DU}{Detection Unit}
\newacronym{ARCA}{ARCA}{Astroparticle Research with Cosmics in the Abyss}
\newacronym{ORCA}{ORCA}{Oscillation Research with Cosmics in the Abyss}
\newacronym{PMT}{PMT}{photomultiplier tube}
\newacronym{CLB}{CLB}{Central Logic Board}
\newacronym{PB}{PB}{Power Board}
\newacronym{IP}{IP}{Intellectual Property}
\newacronym{EMI}{EMI}{Electro-Magnetic Interferences}
\newacronym{WR}{WR}{White Rabbit}
\newacronym{TDC}{TDC}{Time to Digital Converter}
\newacronym{PLL}{PLL}{Phase Locked Loop}
\newacronym{PPS}{PPS}{Pulse Per Second}
\newacronym{SPI}{SPI}{Serial Peripheral Interface}
\newacronym{DC}{DC}{Direct Current}
\newacronym{I2C}{I\textsuperscript{2}C}{Inter Integrated Circuit}
\newacronym{IC}{IC}{Integrated Circuit}
\newacronym{WRPC}{WRPC}{White Rabbit PTP Core}
\newacronym{PTP}{PTP}{Precision Time Protocol}
\newacronym{PCB}{PCB}{Printed Circuit Board}
\newacronym{SyncE}{SyncE}{Synchronous Ethernet}
\newacronym{DRU}{DRU}{Data Recovery Unit}
\newacronym{ToT}{ToT}{Time over Threshold}

\newacronym{HRV}{HRV}{High Rate Veto}

\newacronym{UART}{UART}{Universal Asynchronous Receiver / Transmitter}
\newacronym{RAM}{RAM}{Random Access Memory}
\newacronym{GPS}{GPS}{Global Positioning System}
\newacronym{AES}{AES3}{Audio Engineering Society version 3}
\newacronym{UDP}{UDP}{User Datagram Protocol}

\newacronym{SoC}{SoC}{System on Chip}
\newacronym{LM32}{LM32}{LatticeMico32}
\newacronym{ASIC}{ASIC}{Application Specific Integrated Circuit}
\newacronym{SCL}{SCL}{Serial Clock Line}
\newacronym{FIFO}{FIFO}{First-In First-Out Buffer}
\newacronym{UTC}{UTC}{Coordinated Universal Time} %-- \textit{Temps Universel Coordonn\'e}}
\newacronym{FIT}{FIT}{Failure In Time}
\newacronym{mac}{MAC}{Media Access Controller}
\newacronym{MTTF}{MTTF}{Mean Time To Failure}
\newacronym{HV}{HV}{High Voltage}
\newacronym{OTP}{OTP}{One Time Programmable}
\newacronym{CW}{CW}{Cockroft Walton}

\newacronym{CPLD}{CPLD}{Complex Programmable Logic Device}

\newacronym{EEPROM}{EEPROM}{ Electrically Erasable Programmable Read-Only Memory}
\newacronym{HDL}{HDL}{Hardware Description Language}
\newacronym{SFP}{SFP}{Small Form-Factor Pluggable}
\newacronym{FIDES}{FIDES}{FIDES -- Faith in Latin}
\newacronym{SCB}{SCB}{Signal Collection Board}
\newacronym{FMECA}{FMECA}{Failure Mode, Effects, and Criticality Analysis}
\newacronym{LED}{LED}{Light Emitting Diode}
\newacronym{DAQ}{DAQ}{Data Acquisition System}

\cftpagenumbersoff{figure}
\cftpagenumbersoff{table}

\begin{document}

%\rightmark{15 July 2019}

%\linenumbers 
\maketitle

%\maketitle

%\abstract{}

\begin{abstract}

The KM3NeT research infrastructure being built at the bottom of the Mediterranean Sea will host water-Cherenkov telescopes for the detection of cosmic neutrinos. The neutrino telescopes will consist of large volume three-dimensional grids of optical modules to detect the Cherenkov light from charged particles produced by neutrino-induced interactions. Each optical module houses 31 3-inch photomultiplier tubes, instrumentation for calibration of the  photomultiplier signal and positioning of the optical module and all associated electronics boards. By design, the total electrical power consumption of an optical module has been capped at seven watts.  This paper presents an overview of the front-end and readout electronics system inside the optical module, which has been designed for a 1~ns synchronization between the clocks of all optical modules in the grid during a life time of at least 20 years.

\end{abstract}


% Include email contact information for corresponding author
{\noindent \footnotesize\textbf{*} D. Real,  \linkable{real@ific.uv.es};
\noindent D. Calvo,  \linkable{dacalcia@ific.uv.es}; \noindent  V. van Beveren, \linkable{v.van.beveren@nikhef.nl}
}

\begin{spacing}{2}   % use double spacing for rest of manuscript

% -------------------------------------------------------------------------------------------
\section{Introduction}\label{sec:int}
% -------------------------------------------------------------------------------------------

\begin{figure}[tbp] % figures (and tables) should go top or bottom of
                    % the page where they are first cited or in
                    % subsequent pages
\centering

\begin{subfigure}[b]{.67\linewidth}
\includegraphics[width=1\textwidth]{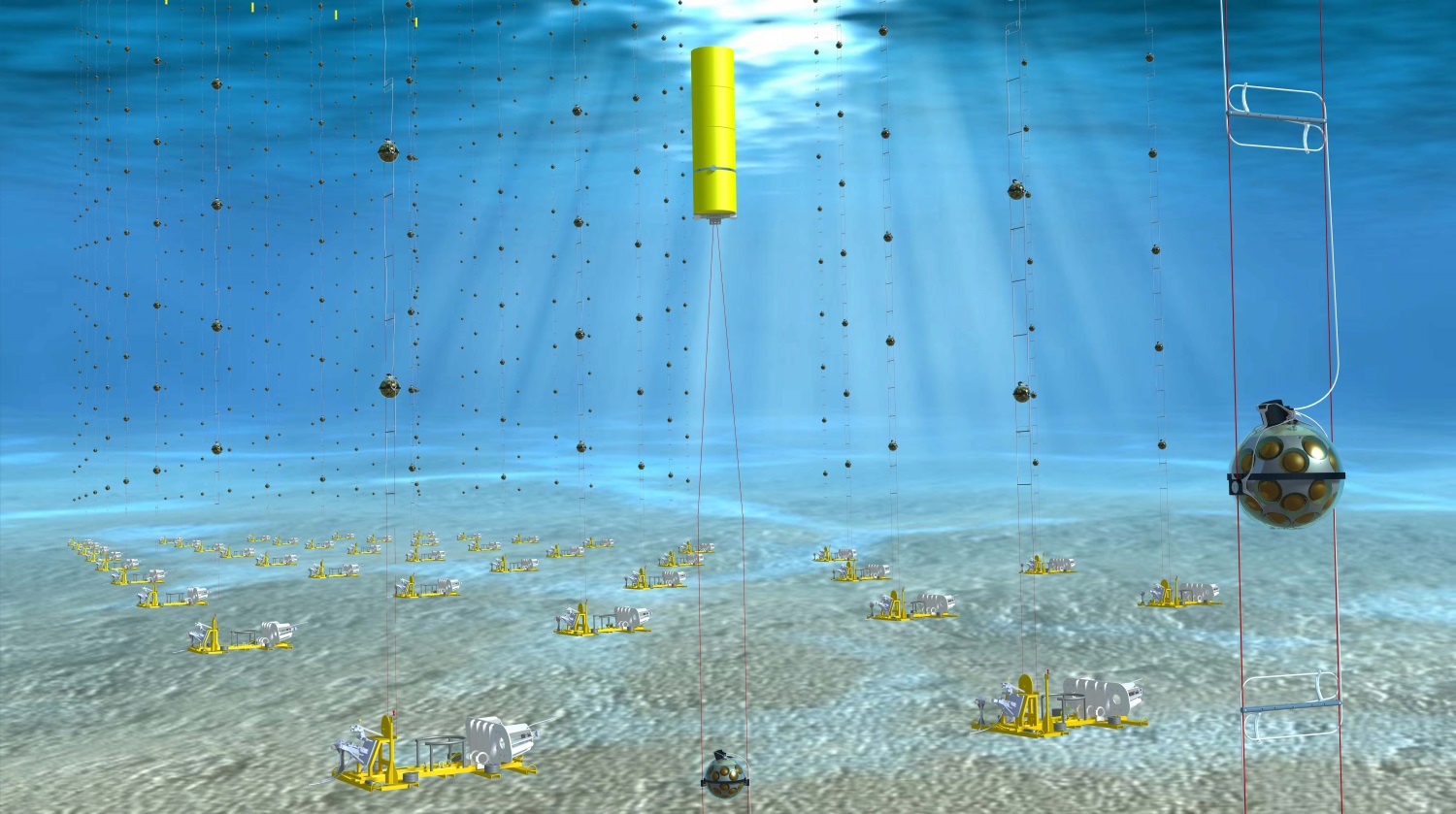}  
\caption{KM3NeT detector, artistic illustration.}
 \label{fig:overview}
\end{subfigure}
\begin{subfigure}[b]{.315\linewidth}
\includegraphics[width=1\textwidth]{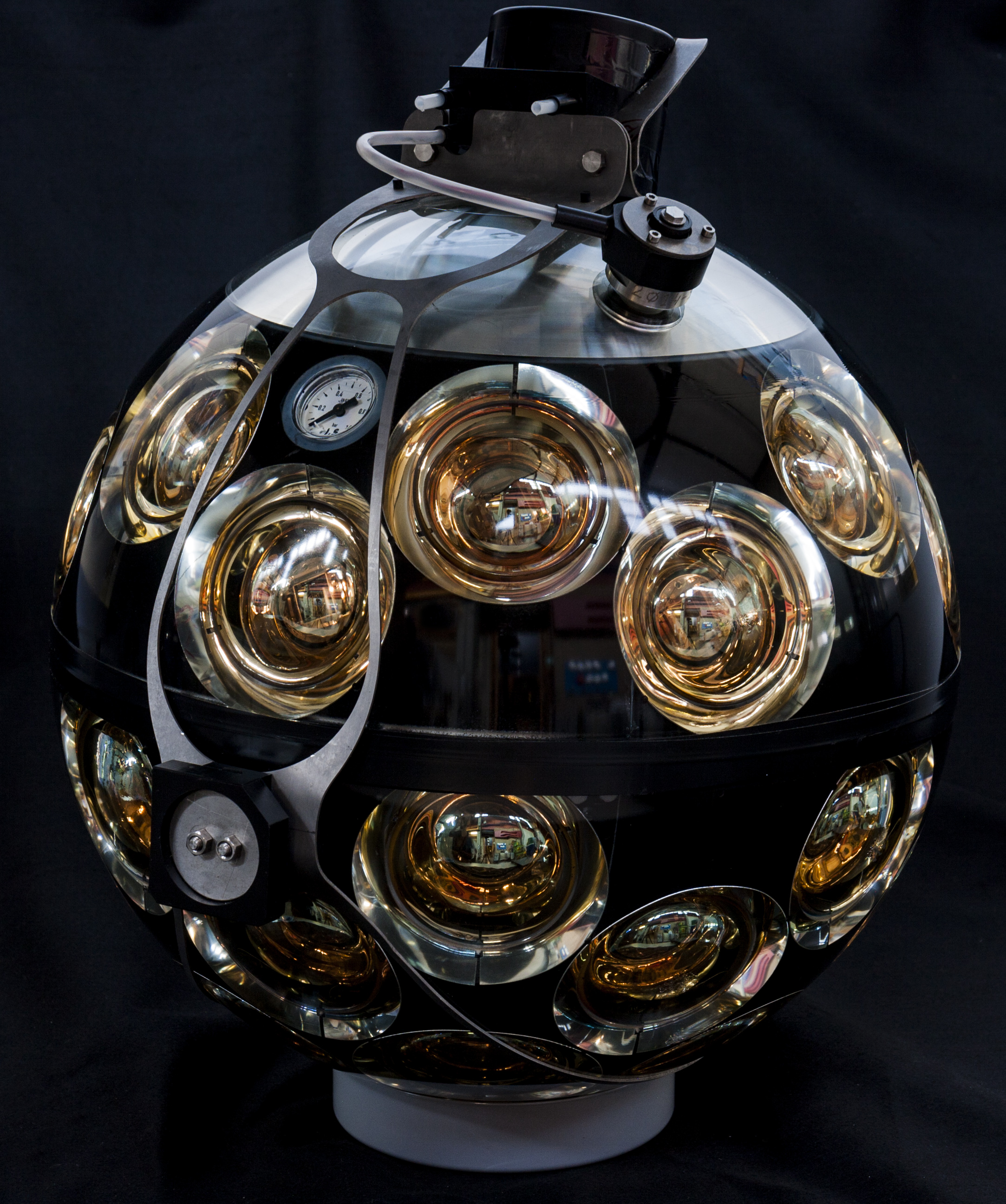}  
\caption{\acrlong{DOM}.}
 \label{fig:domzoom}
\end{subfigure}

\caption{(a) Artistic illustration of the KM3NeT detector. Note that the illustration is not to scale and that actually sunlight will not reach the depths at which the KM3NeT detector is deployed. (b) A picture of the KM3NeT optical module with the flye's eye organisation of the \glspl{PMT} and the cap of the aluminum cooling 'mushroom' visible.  The Titanium collar around the module supports the connection to the ropes of the \gls{DU}.}
\label{fig:dom}
\end{figure}

KM3NeT is a European research facility being built at the bottom of the Mediterranean Sea. It will host the future large volume \acrshort{ARCA} and \acrshort{ORCA} neutrino telescopes. The \acrshort{ARCA} telescope (\acrlong{ARCA}), a cubic kilometer scale detector mainly dedicated to the detection of high energy neutrinos of astrophysical origin, is being installed at a site located offshore the coast of Sicily, Italy, at an approximate depth of 3500 m. The detector of the \acrshort{ORCA} telescope (\acrlong{ORCA}), located at a depth of about 2400 m offshore Toulon, France, will be optimised for the detection of lower energy neutrinos to allow for the study of fundamental properties of neutrinos. \acrshort{ARCA} and \acrshort{ORCA} share the same detector technologies.~\cite{km3net_letter} Cherenkov light produced by neutrino-induced charged particles will be detected by a regular array of optical modules in the water volume of the telescope (Figure~\ref{fig:overview}). Each module (Figure~\ref{fig:domzoom}) is a high-pressure resistant, 17-inch diameter glass sphere containing 31 3-inch \glspl{PMT}, instrumentation for calibration and positioning and all associated electronics boards. The modules are called \glspl{DOM}.~\cite{DOM}\textsuperscript{,}~\cite{DOM2}\textsuperscript{,}~\cite{elec} Eighteen \glspl{DOM}, uniformly distributed along a vertical slender structure, form a \gls{DU}. The \glspl{DOM} are hold into place by means of two thin ropes. The \gls{DU} is anchored on the seabed and kept in a close to vertical position by a submerged buoy at its top. An electro-optical backbone cable, with breakouts at each \gls{DOM}, runs along the full \gls{DU} length providing connection for power feeding and data transmission.

\begin{figure}[tbp] % figures (and tables) should go top or bottom of
                    % the page where they are first cited or in
                    % subsequent pages
\centering

\begin{subfigure}[b]{.6\linewidth}

\includegraphics[trim={1cm 0cm 0cm 0cm}, width=1\textwidth]{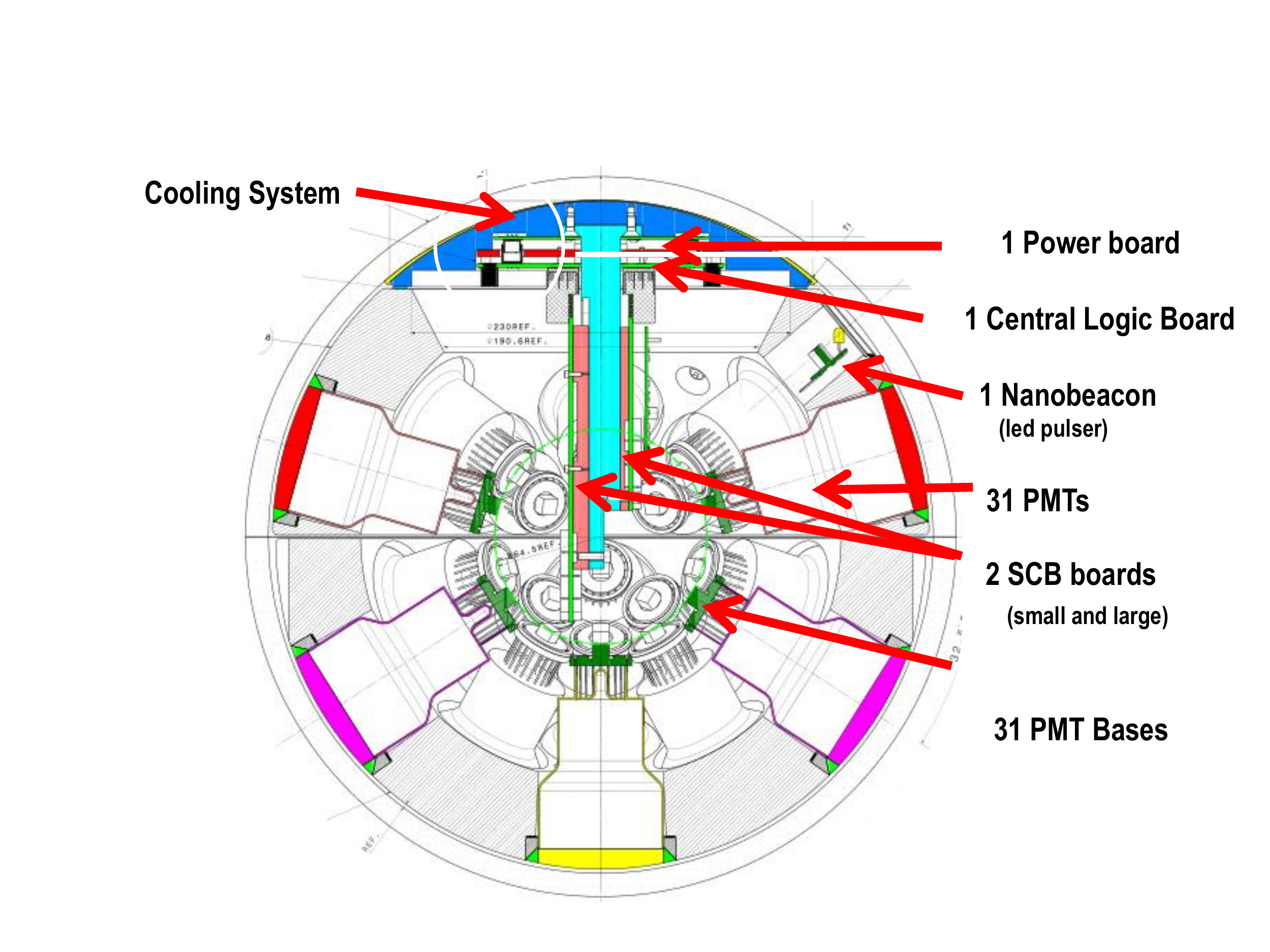}  
\caption{2D vertical cross section of the \gls{DOM}.}
%\label{fig:octb}
\end{subfigure}
\begin{subfigure}[b]{.375\linewidth}
\includegraphics[trim={8cm 0cm 0cm 0cm},clip,width=1\textwidth]{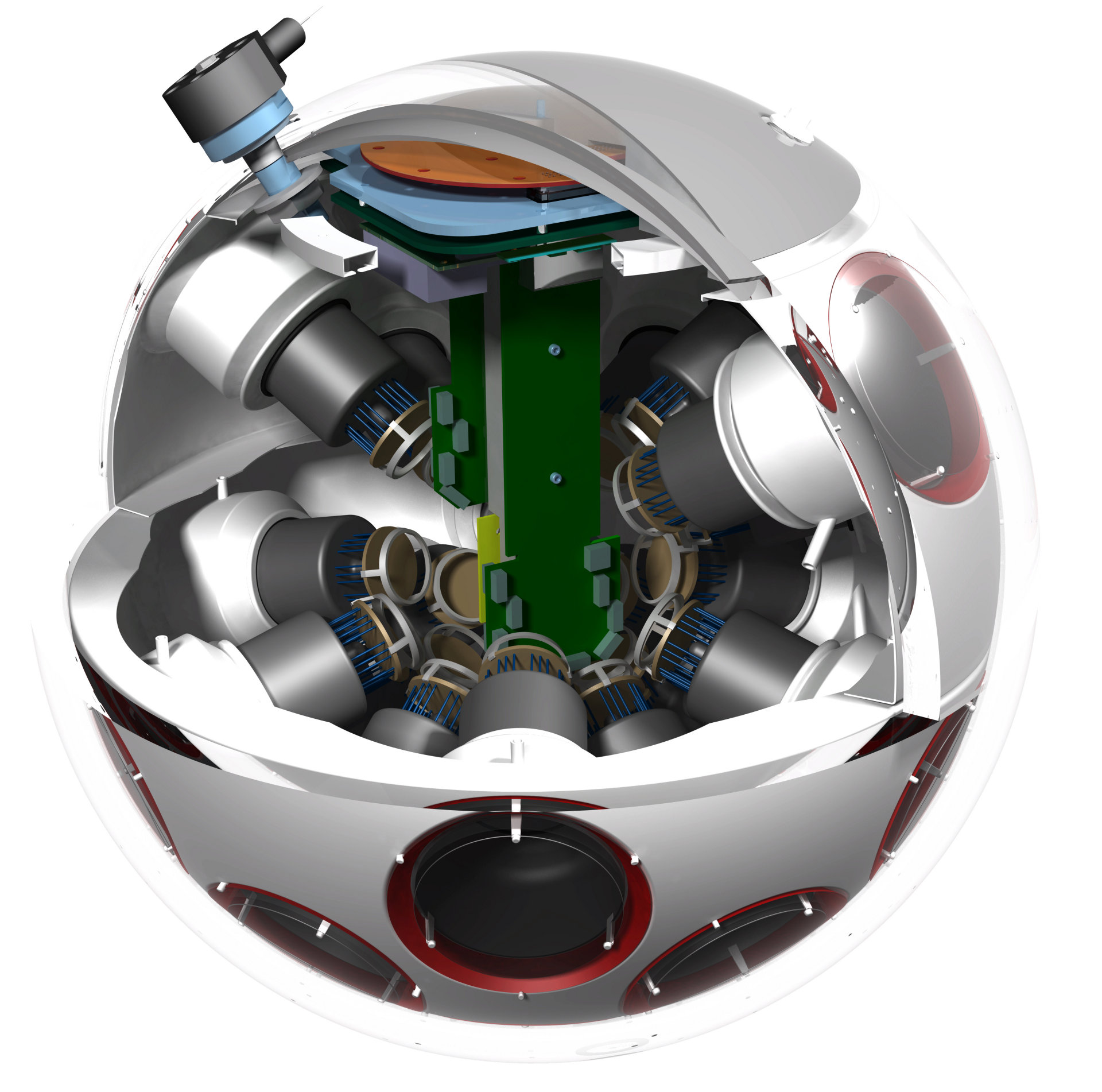}  
\caption{3D open model of the \gls{DOM}.}
%\label{fig:octa}
\end{subfigure}

\caption{2D and 3D drawings of the \gls{DOM} showing the position of the different devices.}
\label{fig:dom2d}
\end{figure}

In each \gls{DOM}, the 31 \glspl{PMT} are organized in five rings of six \glspl{PMT}, plus a single one at the bottom pointing downward (Figure~\ref{fig:dom2d}). The \glspl{PMT} are kept in place by a 3-D printed support structure. The lower and the upper hemisphere of the module contain 19 and 12 \glspl{PMT}, respectively. In the upper hemisphere, a mushroom-shape aluminum structure provides support to the electronic boards of the \gls{DOM}. The top surface of the mushroom cap is glued to the glass sphere in order to provide heat dissipation to the seawater. Fixed to the mushroom cap is the Power Board, which provides all the DC voltages needed by the electronics. This board will be described in Section~\ref{sec:pb}. The \gls{CLB}, which contains a \gls{FPGA}, is directly connected to the Power Board. In the \gls{FPGA}, the \gls{IP} cores that capture the \gls{PMT} generated signals are embedded. Also embedded in the \gls{FPGA} is an implementation of the \gls{WR}~\cite{wr2}, a fully deterministic Ethernet-based timing protocol which provides both data transmission and accurate timing. The \gls{WR} technology allows for a synchronization of the clocks of all \glspl{CLB} in the telescope at nanosecond precision. The description of the \gls{CLB} is presented in Section~\ref{sec:clb}. In Section~\ref{sec:pmtbase}, the \gls{PMT} base board, which generates and adjusts the \gls{HV} supply of the \gls{PMT} and converts the analog signals generated by the \gls{PMT}s to \gls{LVDS} is described. Two \glspl{SCB}, one for each \gls{DOM} hemisphere, connect the \gls{CLB} with the \glspl{PMT} allowing for command and data signal transfer. The \gls{SCB} is described in Section~\ref{sec:oct}.

The light detected by a \gls{PMT} is converted into an electrical pulse. When this electrical pulse surpasses a predetermined threshold, the \gls{PMT} base board sets its \gls{LVDS} output. This output is reset when the electrical pulse goes below the threshold. The first crossing of the threshold and the \gls{ToT} will be measured by the \glspl{TDC} implemented in the \gls{CLB}. The \gls{ToT} gives an estimate of the pulse amplitude and its charge. The calibration of the \gls{PMT} \gls{HV} provides an average \gls{ToT} value of 26.4 ns when a single photoelectron impinges on a \gls{PMT}. The \gls{CLB} organizes the acquisition of the \gls{LVDS} signals in frames, or timeslices, of fixed length in time, typically 100 ms. The data acquired, organized in timeslices, are sent to a computer farm onshore via an optical network integrated in the submarine cables and junction boxes. The \gls{DU} anchor hosts a base module equipped with a wet-mateable jumper to connect the \gls{DU} to the seafloor network. The full chain of readout electronics was successfully qualified in a prototype \gls{DU} of three \glspl{DOM} deployed at the Italian KM3NeT site in May 2014 and operated more than one year~\cite{PRODU}. Mass-produced electronics is operational in full-size deployed \glspl{DU} with 18 \glspl{DOM}~\cite{RATE}, thus demonstrating its reliability.

\begin{figure}[tbp] % figures (and tables) should go top or bottom of
                    % the page where they are first cited or in
                    % subsequent pages
\centering
\includegraphics[width=0.8\textwidth]{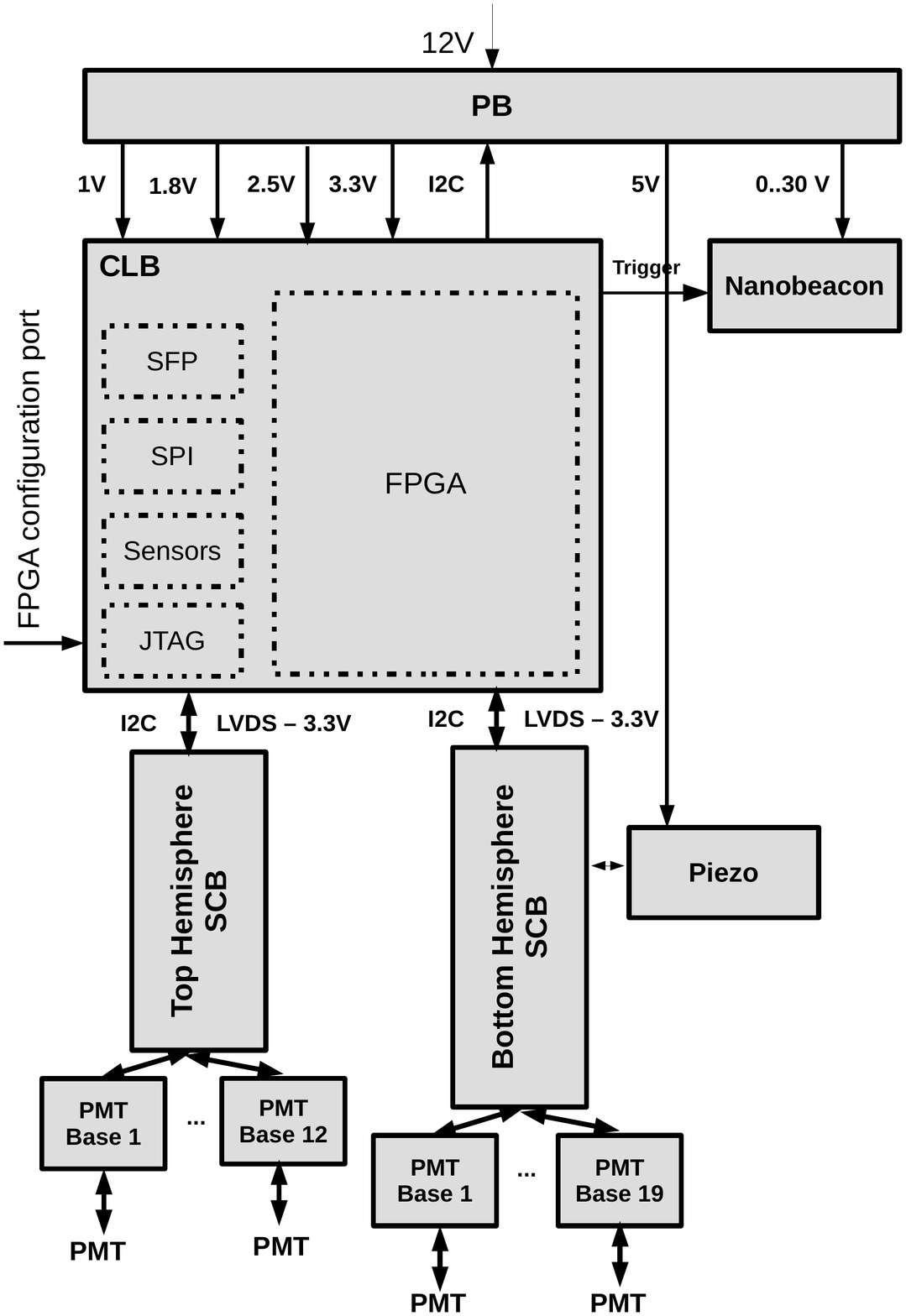}
\caption{Block diagram of the \gls{DOM} electronics boards and their interconnections.}
\label{fig:domfun}
\end{figure}

Figure~\ref{fig:domfun} provides a block diagram of the different \gls{DOM} electronics boards and their interconnections. The power consumption of the \gls{DOM} is discussed in Section~\ref{sec:con} and the reliability studies performed on the \gls{DOM} electronics boards are presented in Section~\ref{sec:fides}. Finally, a conclusion is given about the front-end and readout electronics system in light of the design goals set out by the KM3NeT Collaboration.

% -------------------------------------------------------------------------------------------
\section{The \acrlong{CLB}}\label{sec:clb}
% -------------------------------------------------------------------------------------------

\begin{figure}[tbp] % figures (and tables) should go top or bottom of
                    % the page where they are first cited or in
                    % subsequent pages
\centering
\includegraphics[trim={0cm 2cm 0cm 2cm},clip,width=0.76	\textwidth]{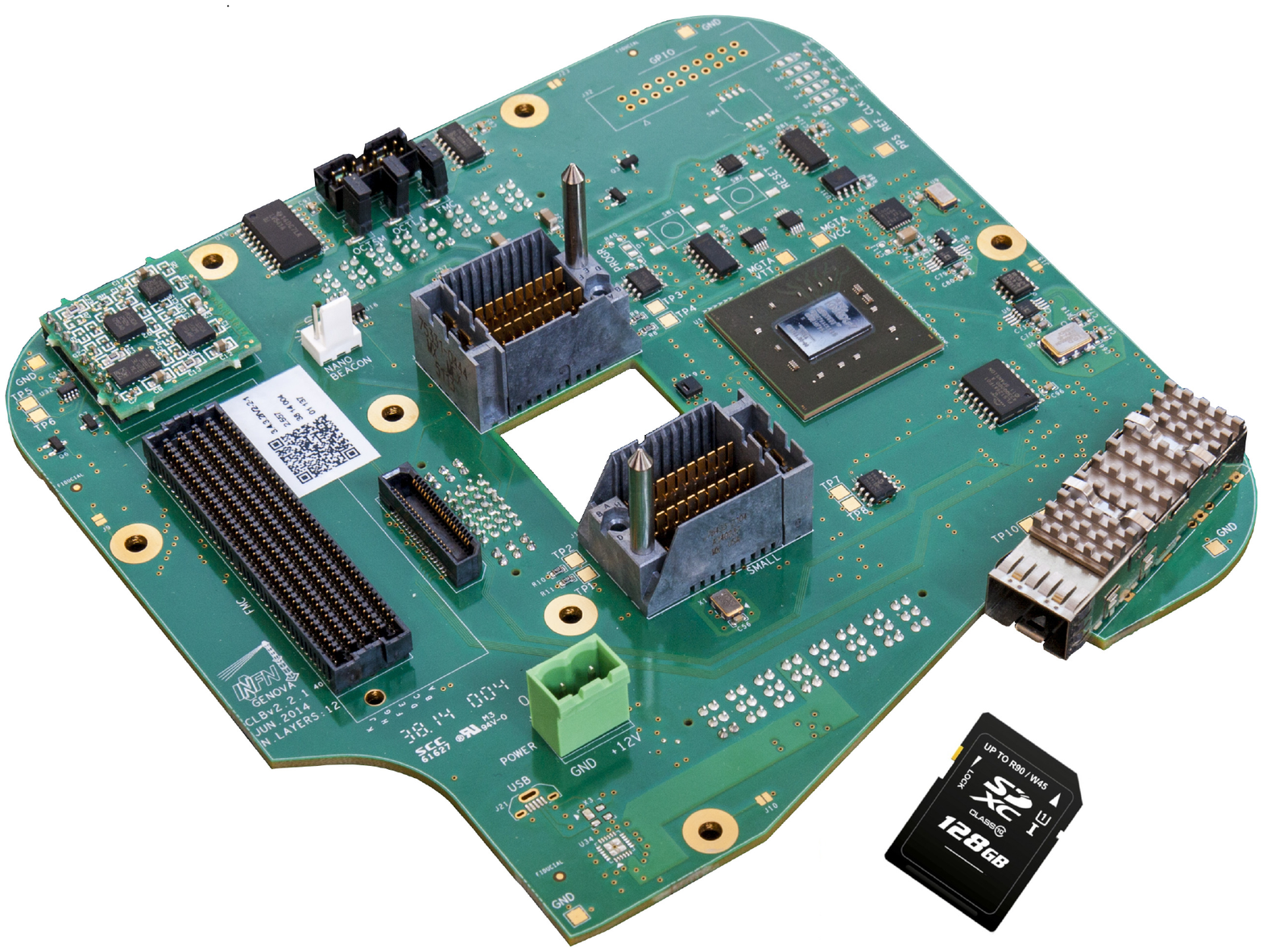}
\caption{ The \gls{DOM} \acrlong{CLB}.  An SD memory is presented close to the \gls{CLB} as dimension reference.}
\label{fig:clbpic}
\end{figure}

The \gls{DOM} \gls{CLB} \cite{clb}\textsuperscript{,}~\cite{clb2}(Figure~\ref{fig:clbpic}) is the main electronics board in the readout chain of KM3NeT. The \gls{PMT} bases generate \gls{LVDS} signals from the \gls{PMT} electrical pulses.  The corresponding \gls{SCB} receives and distributes these signals to the \gls{CLB}, where they are digitized with a resolution of one nanosecond by \glspl{TDC} running in the \gls{FPGA} programmable logic. After being organized and timestamped in the \gls{CLB}, the \gls{TDC} data are transferred to the onshore station for further processing and storage.  The \gls{CLB} board also houses a compass/tiltmeter, three temperature sensors and a humidity sensor. In addition, it provides a connection for a LED flasher, called Nanobeacon~\cite{Nanobeacon}. Also, a piezo sensor is connected to the \gls{CLB} via the \gls{SCB} that serves the lower hemisphere. 

The control of the \gls{CLB} is achieved by means of custom software which runs in a \gls{LM32}~\cite{lm32} soft-processor operating in the programmable logic of the \gls{CLB} \gls{FPGA}.

In the next subsections the hardware, firmware and software of the \gls{CLB} are described.

\begin{figure}[tbp] % figures (and tables) should go top or bottom of
                    % the page where they are first cited or in
                    % subsequent pages
\centering
\includegraphics[width=1.375\textwidth, angle=90]{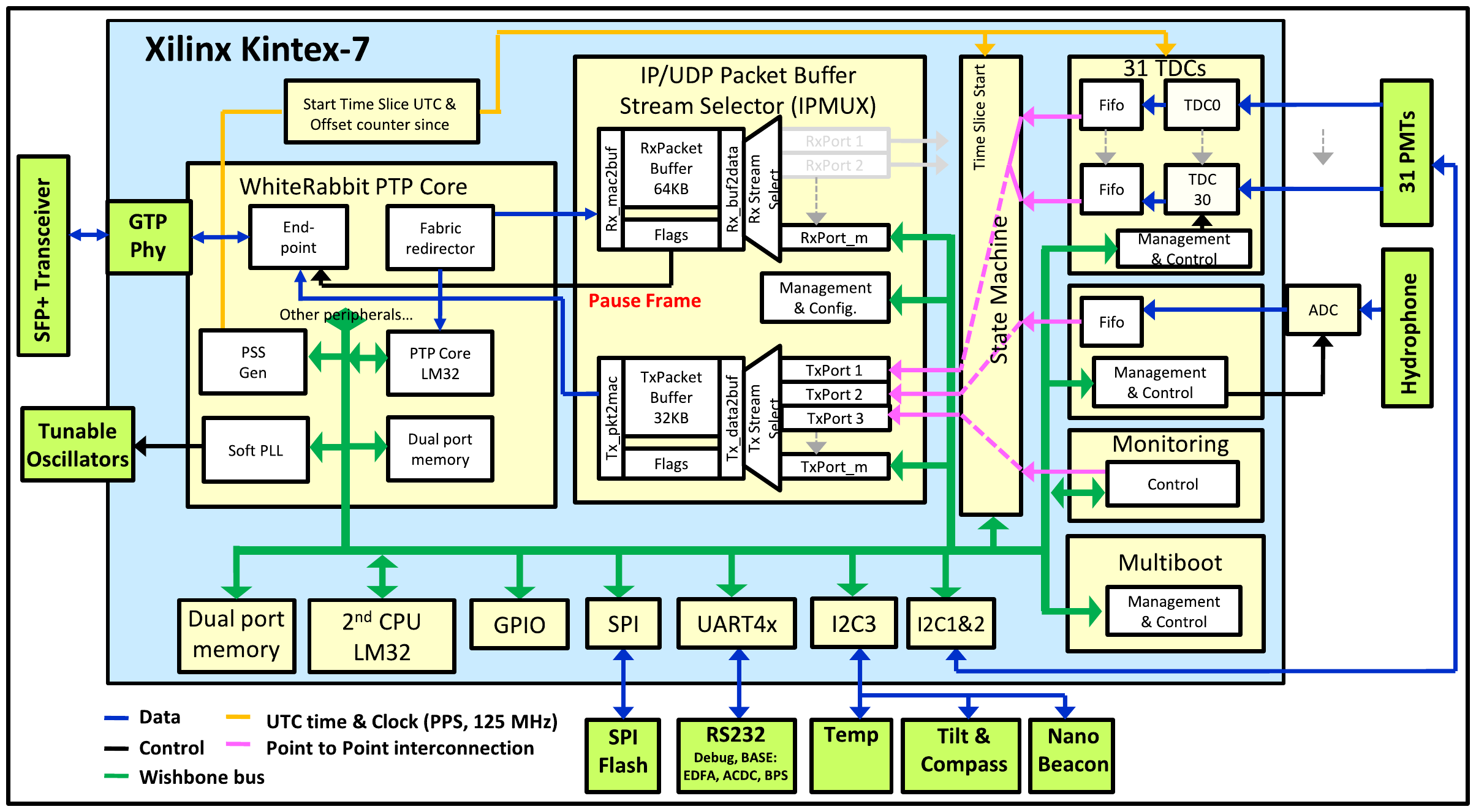}
\caption{Block diagram of the \gls{CLB} \gls{FPGA} firmware.}
\label{fig:clbdia}
\end{figure}

% -------------------------------------------------------------------------------------------
\subsection{\acrlong{CLB} hardware}\label{sec:clbhw}
% -------------------------------------------------------------------------------------------
%The \gls{CLB} hardware consists of the \gls{PCB} and its electrical components. 

The \gls{CLB} \gls{PCB} comprises twelve layers: six of them are dedicated to signals, two of them to power planes while the remaining four layers are ground. The ground layers surround the power planes in order to reduce the \gls{EMI} from the power layers on the signal layers and to improve the signal integrity\cite{Bogatin2009}. For the same reasons, the number of vias in these layers has also been minimized wherever possible.
Special care has been taken in routing the \gls{LVDS} signals generated by the \gls{PMT} base boards. The difference in length between any of the differential pairs has been kept below 100~ps. Moreover, in the case of the clock signals, this difference has been reduced to below 20~ps.

The central coordinating component of the \gls{CLB} is a Xilinx Kintex-7 \gls{FPGA} (XC160-T), chosen for its relatively low power consumption. Other relevant components are: the  \gls{SPI} flash memory, which stores four images of the \gls{FPGA} and the configuration parameters of the \gls{CLB}; the programmable oscillators, which provide the appropriated clock signals needed by the \gls{WR} protocol; two press fit connectors that provide a solid mechanical and electrical connection between the \gls{CLB} and the \gls{SCB}. The \gls{CLB} board includes a 25~MHz crystal oscillator. The oscillator signal is first transferred from a clock pin to a buffer in the \gls{FPGA}, and then fanned-out to the inner \gls{PLL} to provide two high frequency clocks of 250~MHz but with a $90^{\circ}{\rm }$ phase shift, needed by the \gls{TDC} core. The main component used for the communications with the onshore station is the \gls{SFP} transceiver, which interfaces the electronics with the optics system.

% -------------------------------------------------------------------------------------------
\subsection{\gls{CLB} firmware}\label{sec:clbfw}
% -------------------------------------------------------------------------------------------
The readout logic of the \gls{DOM} runs in the programmable fabric of the \gls{FPGA}. A block diagram of the readout logic is shown in Figure~\ref{fig:clbdia}. Its main blocks are:

\begin{itemize}
\item The \gls{LM32} soft-processor running the control and monitoring software for the \gls{CLB}.
\item The \gls{WRPC} (\gls{PTP}), which implements the \gls{WR} protocol.
\item The \glspl{TDC}, which digitize and time-stamp the \gls{PMT} signals arriving at the \gls{CLB}.
\item The state machine and IPMux cores, which collect the \gls{TDC} data from the \glspl{PMT}, \gls{AES} data from the piezo sensor and the monitoring data from the \gls{LM32}, and dispatch them over Ethernet to the onshore station~\cite{daq}.
\item The multiboot core, which allows safe remote reconfiguration of the \gls{FPGA} firmware.
\item The different control cores for the instrumentation.
\end{itemize}

\begin{figure}[tbp] % figures (and tables) should go top or bottom of
                    % the page where they are first cited or in
                    % subsequent pages
\centering
\includegraphics[trim={2.5cm 0cm 0cm 0},clip,width=1.1\textwidth]{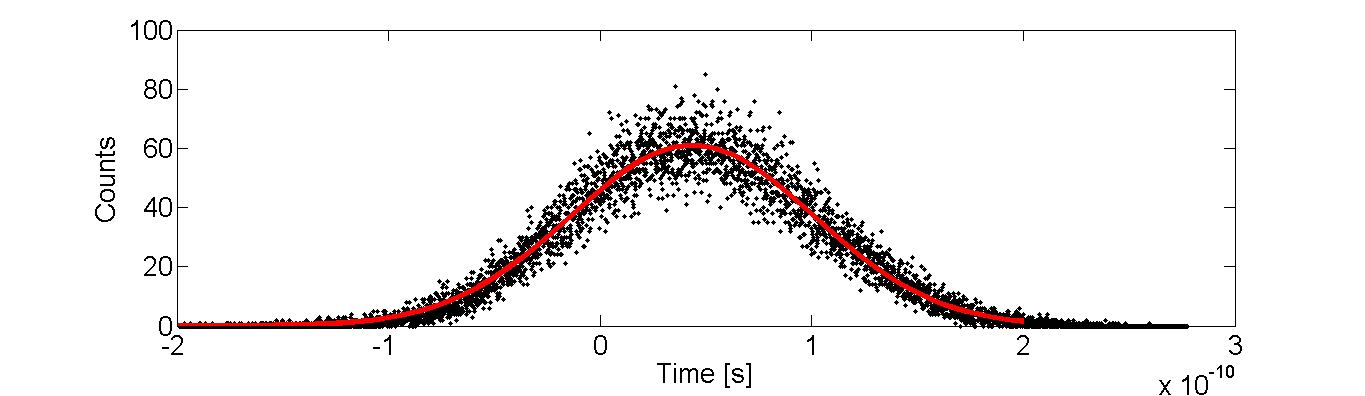}
\newline 
\caption{Skew, measured at the laboratory, of the \acrlong{PPS} of the \gls{CLB}  with respect to the \acrshort{PPS} of the Master White Rabbit Switch. The red line is a Gaussian fit to the data with 22 ps standard deviation.}
\label{fig:wr}
\end{figure}

\begin{figure}[bp] % figures (and tables) should go top or bottom of
                    % the page where they are first cited or in
                    % subsequent pages
\centering
\includegraphics[trim={1cm 0cm 0cm 0},clip,width=1.11\textwidth]{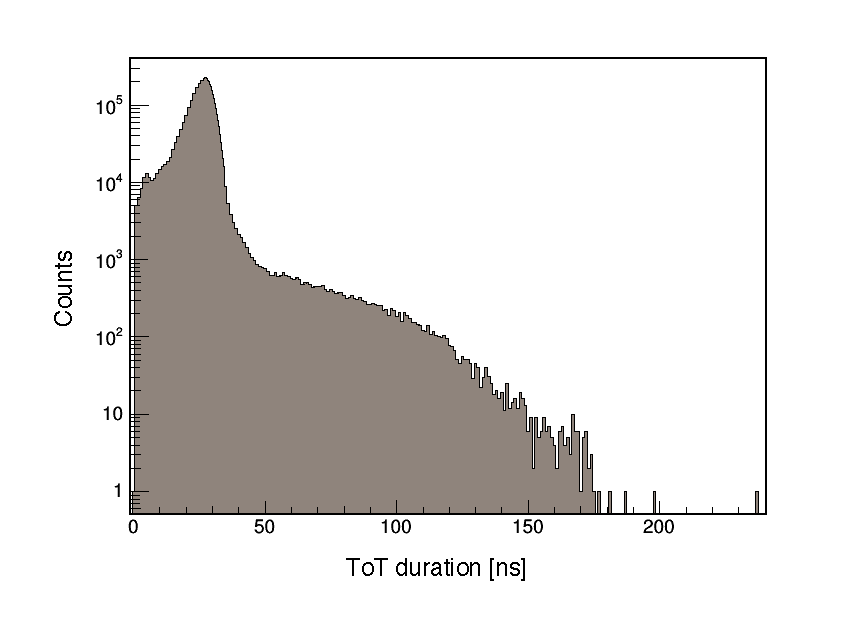}
\caption{\Gls{ToT} distribution from one \gls{DOM} of a KM3NeT \gls{DU}. All the channels have been calibrated in order to harmonise the \gls{ToT} data arriving from different \glspl{PMT}}
\label{fig:tot}
\end{figure}

% -------------------------------------------------------------------------------------------
\subsubsection{Soft-Microcontroller}\label{sec:soc}
% -------------------------------------------------------------------------------------------

Central to the control and monitoring of the \gls{CLB} is the \gls{LM32}, a subsection of the \gls{FPGA} fabric consisting of a \gls{CPU}, \gls{RAM} and peripherals for timing and communication (\gls{UART}, \gls{SPI} and \gls{I2C}). The \gls{LM32} was chosen because it uses less \gls{FPGA} resources than other \glspl{CPU}~\cite{softcpustudy} and has a wishbone bus~\cite{wish} master interface. For the wishbone bus many open-source programmable logic peripherals exist, such as \gls{SPI}/\gls{I2C} controllers, co-processors, timers and counters. In addition, the \gls{LM32} \gls{CPU} is also used in the \gls{WRPC}, easing the integration and reducing the design complexity. The \gls{CPU} runs at 62.5~MHz, and has 128~kB of combined program and data \gls{RAM}. Moreover, the wishbone bus also connects to the KM3NeT specific programmable logic cores, such that the \gls{LM32} can control and monitor the peripherals, which are discussed in the next sections.

% -------------------------------------------------------------------------------------------
\subsubsection{\acrlong{WR} \acrshort{PTP} core}\label{sec:wrptpcore}
% -------------------------------------------------------------------------------------------
The \gls{WRPC} is an enhanced Ethernet \gls{mac}, embedded in the \gls{CLB} \gls{FPGA} programmable logic. Apart from transferring data, as a regular Ethernet \gls{mac} does, the \gls{WR} protocol synchronizes all \gls{CLB} clocks in the detector. The protocol is based on the \gls{SyncE} and \gls{PTP} standards.~\cite{wr1} The \gls{WRPC} synchronizes the \gls{CLB} through the same optical link that is used for data transmission. The global time of the \gls{WR} network is provided by the \gls{WR} master switch located onshore, which is synchronized to a \gls{GPS} receiver.  The \gls{WRPC} \gls{IP} core synchronizes with the \gls{WR} master switch and provides a register with the \gls{UTC}, which is used by the rest of the \gls{CLB} firmware. It also outputs a \gls{PPS} signal, which rising edge occurs precisely at the second transition. In order to qualify the stability of the synchronization at the \gls{CLB}, the  skew between the \gls{PPS} of the \gls{CLB} and the \gls{PPS} of a \gls{WR} switch has been measured at the laboratory using a 50 km optical fiber connection. The skew has a Gaussian distribution with 22~ps standard deviation (Figure~\ref{fig:wr}). 

% -------------------------------------------------------------------------------------------
\subsubsection{State machine}\label{sec:stmach}
% -------------------------------------------------------------------------------------------
The data acquisition is organized in consecutive frames with a period of typically 100~ms, called timeslices. The state machine core orchestrates the data acquisition for the \gls{CLB}. Firstly, it is responsible for generating the periodic start of the timeslice signal. This signal is synchronized to the start of a \gls{UTC} second and repeat at the start of every period. All data acquiring \gls{IP} cores synchronize their acquisition to this start timeslice signal, and all acquired data are sectioned and timestamped relative to it.  Secondly, the state machine is responsible for gathering the acquired data, and merging the \gls{UTC} time of the timeslice start signal, called the super time, to the acquired data. By combining the relative timeslice time and the super time, the \gls{UTC} time for all acquired data can be resolved by the onshore \gls{DAQ}. Once the acquired data are ready, the last responsibility of the state machine is to package the data to be dispatched towards the \gls{UDP} packet generator (IPMux). The data are portioned into frames such that they will fit within the payload of a \gls{UDP} jumbo packet. A frame header containing metadata, such as the stream identifier or the run number, is also prepared.

% -------------------------------------------------------------------------------------------
\subsubsection{\acrlong{TDC}}\label{sec:tdc}
% -------------------------------------------------------------------------------------------
The \glspl{TDC} sample the signals from the \gls{PMT} bases. They are implemented in the \gls{CLB} \gls{FPGA} programmable logic with one \gls{TDC} channel per \gls{PMT}, totalling 31 \gls{IP} cores. The cores measure both the pulse arrival time and the duration of the pulse (\gls{ToT}) using the 1~ns precise \gls{UTC} \gls{WRPC} time. The distribution of the \gls{ToT} data readout as measured by one \gls{DOM} is shown in Figure~\ref{fig:tot}. The \gls{TDC} core produces 48~bits per event, where the first eight most significant bits are used for the \gls{PMT} identifier, the next 32 bits code the arrival time of the event with respect to the timeslice start time, and the last eight bits code the duration. The events are then dispatched to the state machine which also organizes the \gls{TDC} acquisition in timeslices.

\begin{figure}[tbp] % figures (and tables) should go top or bottom of
                    % the page where they are first cited or in
                    % subsequent pages
\centering
\includegraphics[trim={0cm 3.5cm 0cm 2cm},clip,width=1.1\textwidth]{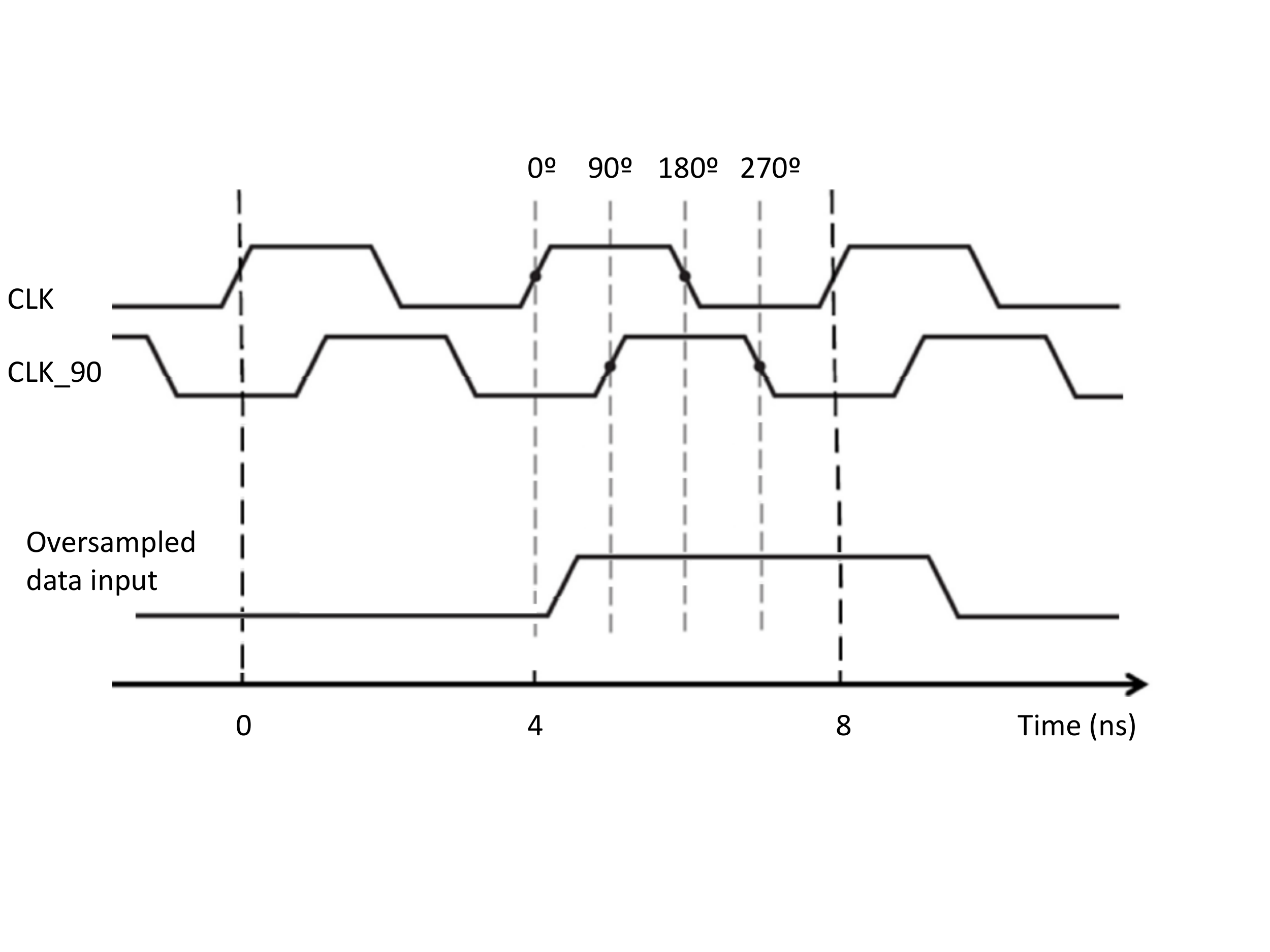}
\caption{Oversampling technique using two phase-shifted clocks. Technique implemented in the \gls{CLB} \glspl{TDC}.}
\label{fig:clbover}
\end{figure} 

The system clock of the \gls{FPGA} firmware is derived from the 25~MHz hardware quartz in the \gls{PCB}. This clock signal is first transferred to a digital \gls{PLL} to generate the system frequency of 62.5~MHz. The White Rabbit protocol adjusts the phase and frequency of the \gls{FPGA} system clock to the reference master clock. Finally, the adjusted system clock is fanned out to the inner \glspl{PLL} within the \gls{FPGA} to provide two high frequency clocks of 250~MHz with a phase shift of $90^{\circ}$. The \gls{TDC} input signals are oversampled at 1 ns rate using the rising and falling edge of the two clocks of 250~MHz as shown in Figure~\ref{fig:clbover}.

The information of the sampling is organized by the \glspl{TDC}, where the arrival time and the pulse length are coded. As all other \gls{CLB} \gls{IP} cores connected to the \gls{LM32} soft \gls{CPU}, the \gls{TDC} core is controlled by the \gls{LM32} itself, allowing for enabling/disabling any of the 31 \gls{TDC} channels.

The \glspl{TDC} implement the \gls{HRV} and Multihit features. The \gls{HRV} limits the total number of acquired hits in a timeslice. If the number of events in a \gls{TDC} channel surpasses a predetermined threshold, the acquisition is stopped in that channel until the start of the next timeslice. In this way, it is possible to limit the amount of data sent onshore, preventing the blockage of the data acquisition.
The Multihit option allows to expand the range of the \glspl{TDC}, limited by the \gls{ToT} codification of eight bits. If this option is active, then the hits with a \gls{ToT} longer than 255 ns are coded as two or more consecutive events.

% -------------------------------------------------------------------------------------------
\subsubsection{Acoustic readout}\label{sec:acrd}
% -------------------------------------------------------------------------------------------

The \gls{CLB} also includes a core for the readout of the acoustic piezo sensor~\cite{piezo}, one of the positioning instrumentation devices installed in the \gls{DOM}. This core reads out the acoustic piezo channel data and timestamps it with respect to the \gls{WRPC} time. In addition, it generates from the raw acoustic data an \gls{AES} formatted stream, which is dispatched to the state machine.

% -------------------------------------------------------------------------------------------
\subsubsection{IPMux}\label{sec:ipmux}
% -------------------------------------------------------------------------------------------
The packets created by the state machine are sent to one of the input ports of the IPMux, an IP/UDP packet buffer stream selector. The IPMux has different input ports for each data sources. For each packet received from the state machine a \gls{UDP} header is added. By using the Ethernet jumbo frames, a maximum transfer unit of 9014 bytes per frame is possible, consequently reducing protocol overhead significantly when the channel is fully occupied.

The IPMux receives also data from \glspl{TDC}, the acoustic readout, the monitoring and the slow control \gls{LM32} channels. All of them are aggregated on the IPMux and transferred to the \gls{WRPC} endpoint, where they are routed through the White Rabbit core and sent onshore. Once they arrive to shore, it is possible to discriminate any of the sources of the IPMux (optical, acoustic and monitoring channel) by the port number.

% -------------------------------------------------------------------------------------------
\subsubsection{Monitoring channel}\label{sec:monchan}
% -------------------------------------------------------------------------------------------
The monitoring channel enables transmission of metadata synchronous with the \gls{TDC} and \gls{AES} channels. However, unlike the \gls{TDC} and \gls{AES}, the monitoring channel is not data driven, and produces only one packet of content at the timeslice start signal. The header of the packet provides information regarding the \gls{TDC} \gls{FIFO}. The monitoring packet consists of two parts. The first part is delivered by programmable logic, containing additional summary information concerning the \gls{TDC} channel, such as the actual number of hits per channel. The content of the second part is software defined. At initialization, the programmable logic is provided with a pointer to a software defined structure. For each timelice, the content of this structure is combined before dispatching to the state machine.

The software provides additional information such as the latest reading from the compass and tilt sensor. Also information about the state of the buffers and other system information is inserted into this packet.

% -------------------------------------------------------------------------------------------
\subsubsection{Multiboot core}\label{sec:rrs}
% -------------------------------------------------------------------------------------------
On startup, the \gls{FPGA} configures itself by loading the first valid image it finds while scanning the SPI flash memory. Up to four images can be stored in the flash memory at subsequent memory locations, reserving the memory regions above those images for storage of settings and logging. The multiboot gives access to the internal Xilinx specific ICAP2 hard-IP block, which allows software initiated reconfiguration of the FPGA at any memory offset. The multiboot is an essential part of the two-stage startup sequence used for fail-safe startup of the CLB. The multiboot mechanism is described in \ref{sec:multiboot}.

% -------------------------------------------------------------------------------------------
\subsection{CLB embedded software}\label{sec:embsw}
% -------------------------------------------------------------------------------------------
The \gls{FPGA} contains two \gls{LM32} processors, the \gls{WRPC} \gls{LM32} and the second \gls{LM32}. Both run a separate software stack. The \gls{WRPC} \gls{LM32} software was developed by the White Rabbit Collaboration,~\cite{broadcast} but it has been adapted to the KM3NeT network topology. The second \gls{LM32} controls the \gls{DOM}. The software has been developed by the KM3NeT Collaboration from scratch and designed as control software for the KM3NeT detector. The latter software is discussed in the following sections.

\begin{figure}[tbp] % figures (and tables) should go top or bottom of
% the page where they are first cited or in
% subsequent pages
\centering\includegraphics[trim={0cm 22cm 16cm 1cm},clip,width=1\textwidth]{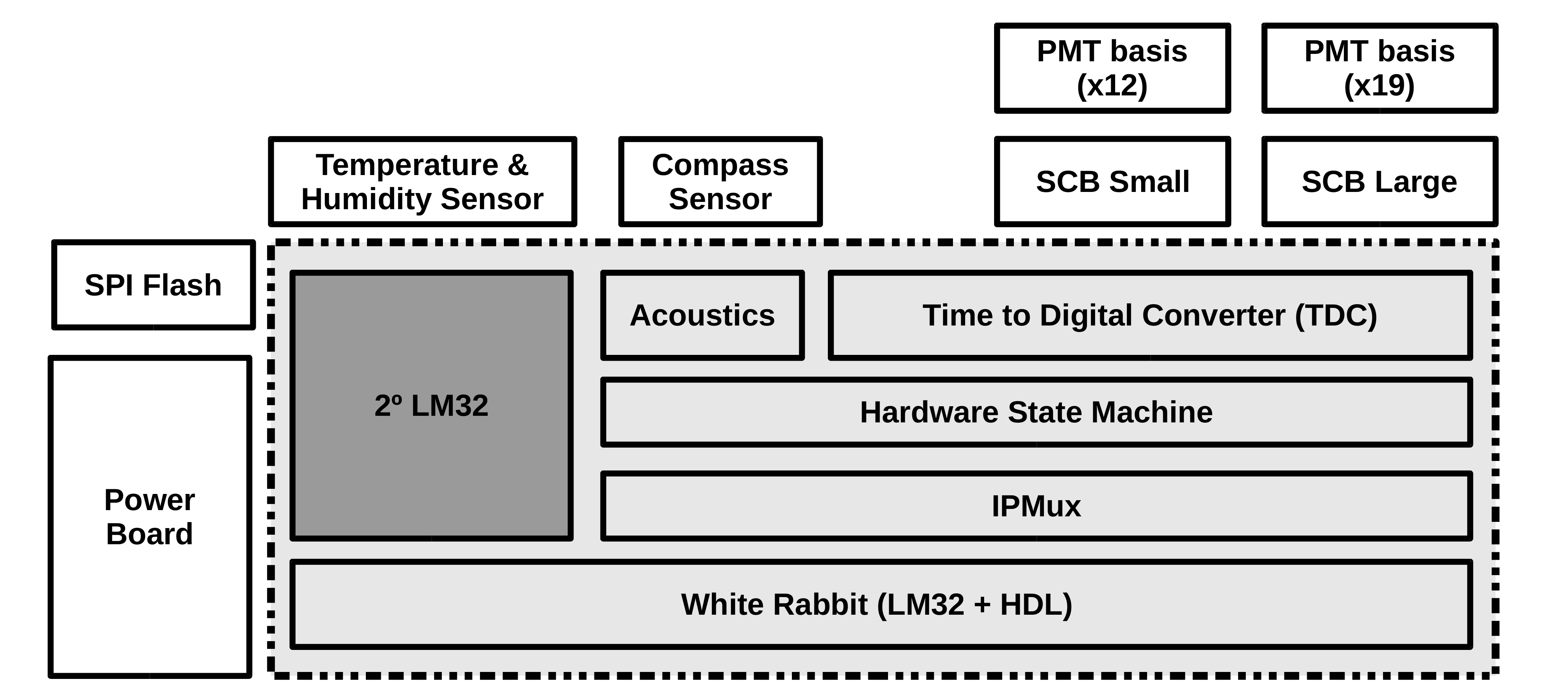}
\newline
\caption{Embedded software location with respect to the main components of the \gls{CLB}, shown in gray. The dashed line shows the boundaries of the \gls{FPGA}.}
\label{fig:softhard}
\end{figure}

\begin{figure}[tbp] % figures (and tables) should go top or bottom of
                    % the page where they are first cited or in
                    % subsequent pages
\centering
\includegraphics[trim={0.5cm 3cm 0cm 0},clip,width=.95\textwidth]{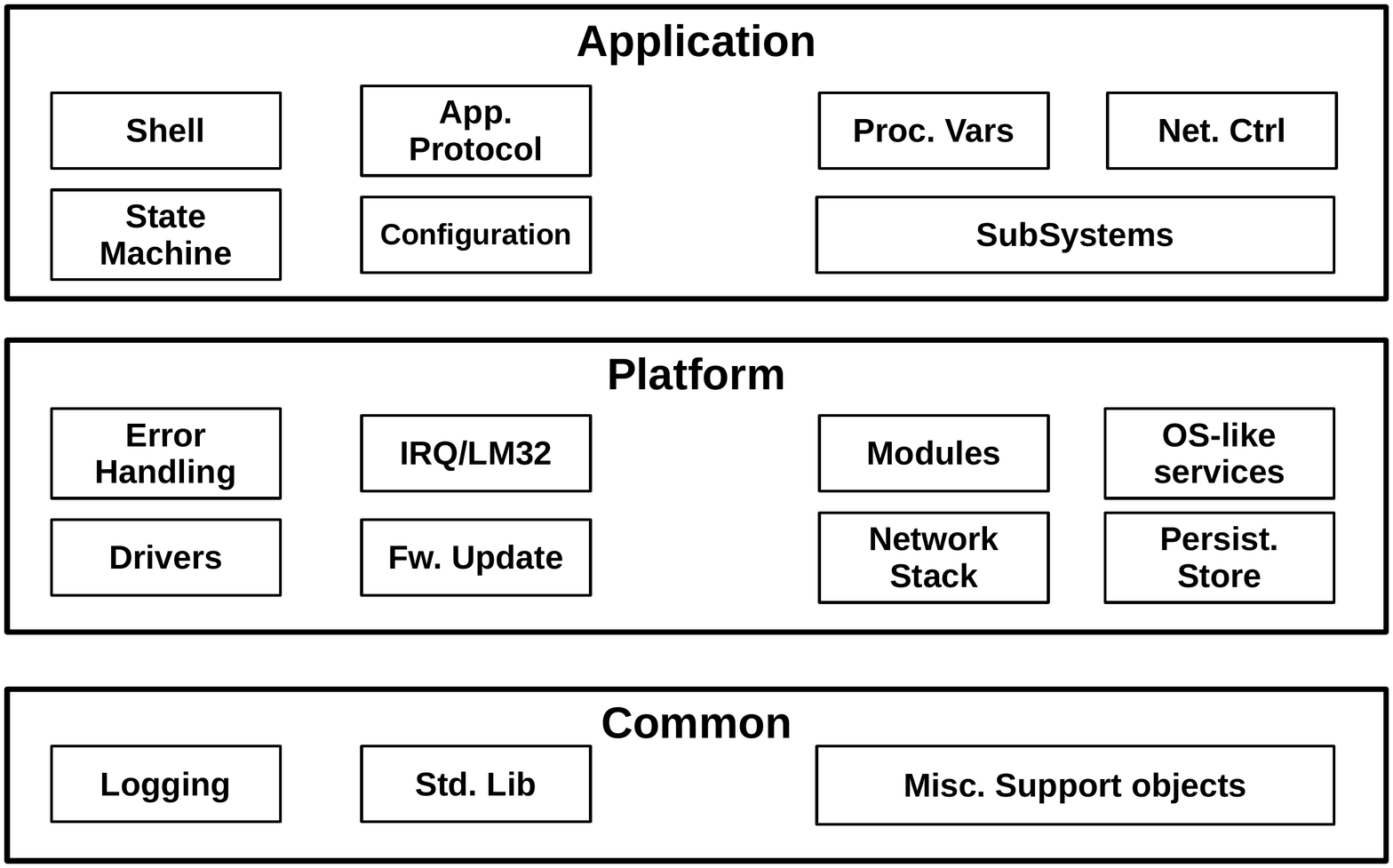}
\newline
\caption{Layers and modules of the \gls{CLB} embedded software.}
\label{fig:softstack}
\end{figure}

% -------------------------------------------------------------------------------------------
\subsubsection{Main tasks}\label{sec:emswtask}
% -------------------------------------------------------------------------------------------
The KM3NeT embedded software handles the following tasks: 

\begin{itemize}
\item Initializing, controlling and monitoring hardware.
\item Executing commands issued from the onshore station.
\item Sending diagnostic information back to shore.
\item Applying firmware updates.
\end{itemize}

A representation of the hardware directly coupled to the second \gls{LM32} is shown in Figure~\ref{fig:softhard}. Most of the components inside the dashed line are programmable logic cores, including the \gls{CPU}s. The hardware devices lie outside the dashed line boundary. Almost all cores are mapped into the memory space of the \gls{LM32}. The embedded software reads from or writes to specific memory locations, depending on the device addressed. Outside the dashed lines in Figure~\ref{fig:softhard}, the hardware is controlled through  external \gls{IC} buses like \gls{I2C} or \gls{SPI}. The interfaces require an additional layer of drivers to communicate with these devices.

% -------------------------------------------------------------------------------------------
\subsubsection{Software generalities}\label{sec:software}
% -------------------------------------------------------------------------------------------
The software running in the \gls{CPU} is primarily coded in C, but some bootstrap and interrupt handling code has been written in \gls{LM32} assembly. There is no preemptive embedded operating system, but just a simple kernel capable of executing different tasks in a collaborative fashion. The layered structure of the main modules of the software stack is shown in Figure~\ref{fig:softstack}.

The Common layer contains functions and objects used throughout the software. They are not specific to the \gls{LM32} platform and could be compiled for any architecture. For example, the logging facilities are placed in this layer.
The Platform layer contains all the code required to control the \gls{LM32} and all the connected hardware. It consists of hardware control, OS-like services and the network stack. It does not contain application functionality, but it provides convenient functions for the application layer to control and monitor the hardware.
Finally, the Application layer contains the high-level functionality of the software. It interprets and executes remote commands, configures and monitors the hardware and implements the software state machine.

% -------------------------------------------------------------------------------------------
\subsubsection{Firmware update and multiboot}\label{sec:multiboot}
\begin{figure}[tbp] % figures (and tables) should go top or bottom of

\centering
 \includegraphics[width=.55\textwidth]{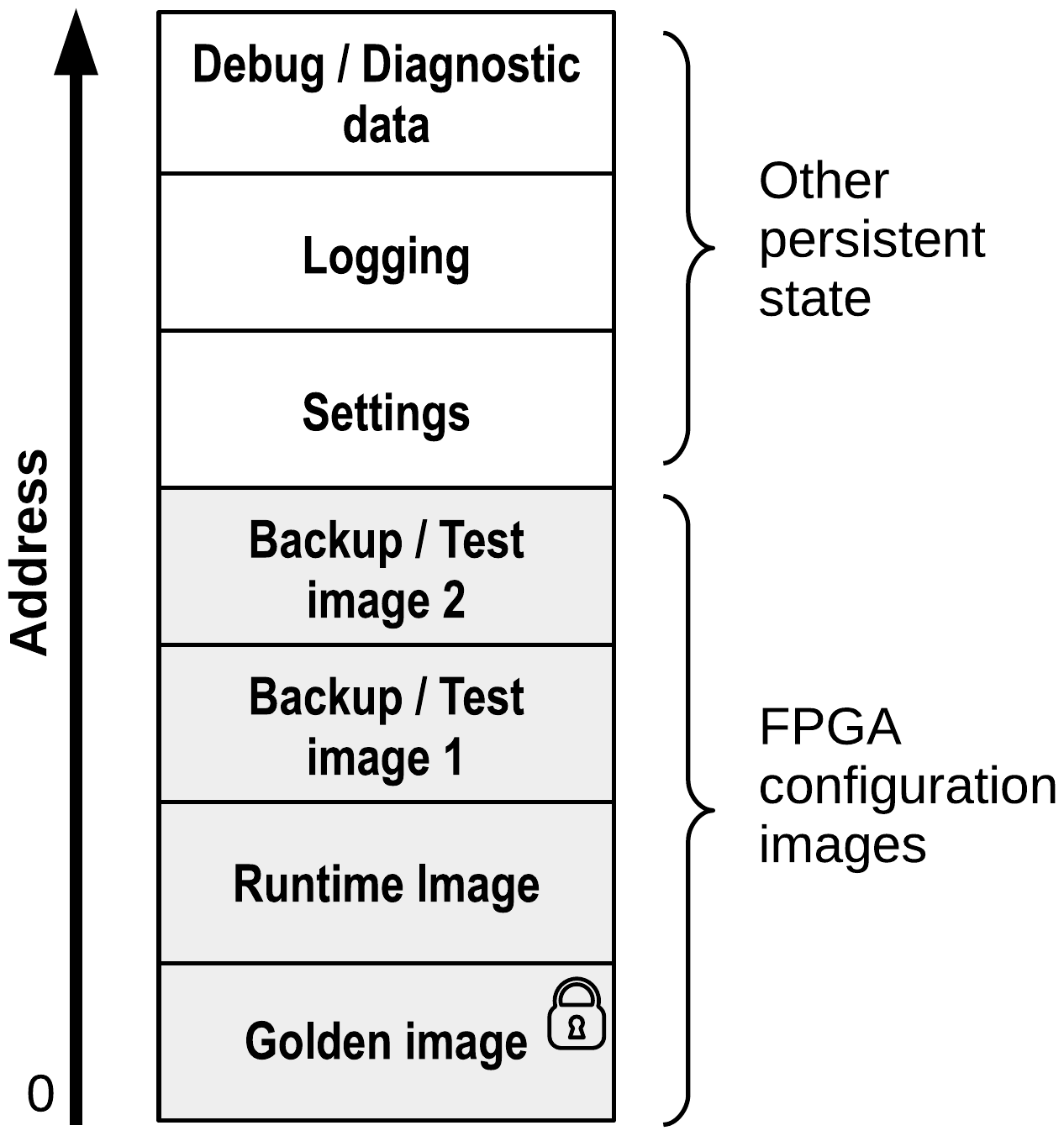}
 \vspace{5mm}
\caption{Content of the serial flash memory. At address 0 starts the golden startup image, followed by the primary runtime image, and two backup images. The remainder of the memory contains various types of persistent states.}
\label{fig:flash}
\end{figure}

% -------------------------------------------------------------------------------------------
As explained in subsection~\ref{sec:rrs}, the embedded software has, through the multiboot core, the capability of configuring the \gls{FPGA} from any image located in the serial flash. For the \gls{CLB}, the flash may contain up to four separate configuration images, starting at address 0 with the startup image, also known as the golden image. The subsequent image is the runtime image, then follow two possible backup images, or test images. After this, the space is reserved for settings and persistent logging. The remaining area of the flash is reserved for storing custom debug and diagnostic information. The complete flash layout is shown in Figure~\ref{fig:flash}. 

The golden image is a special image with minimal hardware initialization. The memory region occupied by the golden image is protected from accidental overwrite by write protection feature part of the flash controller. The golden image will start by default a pre-selected image, usually the runtime image, 30~s after a network connection has been established. In exceptional conditions, the startup procedure can be aborted from shore in the 30~s window. The golden image also provides access to diagnostic and recovery features. Each image on the flash can be updated by remote, including the golden image. However, the latter is an exceptional case and should be avoided. To deal with such cases, a precise and safe procedure has been prepared, which safeguards against accidental loss of \gls{CLB} due to lack of valid images in the flash. The procedure requires that at least one valid image is always present in the serial flash, even during update.

% -------------------------------------------------------------------------------------------
\section{Power Board}\label{sec:pb}
\begin{figure}[tbp] % figures (and tables) should go top or bottom of
                    % the page where they are first cited or in
                    % subsequent pages
\centering
\includegraphics[trim={0cm 0cm 0cm 0cm},clip,width=1\textwidth]{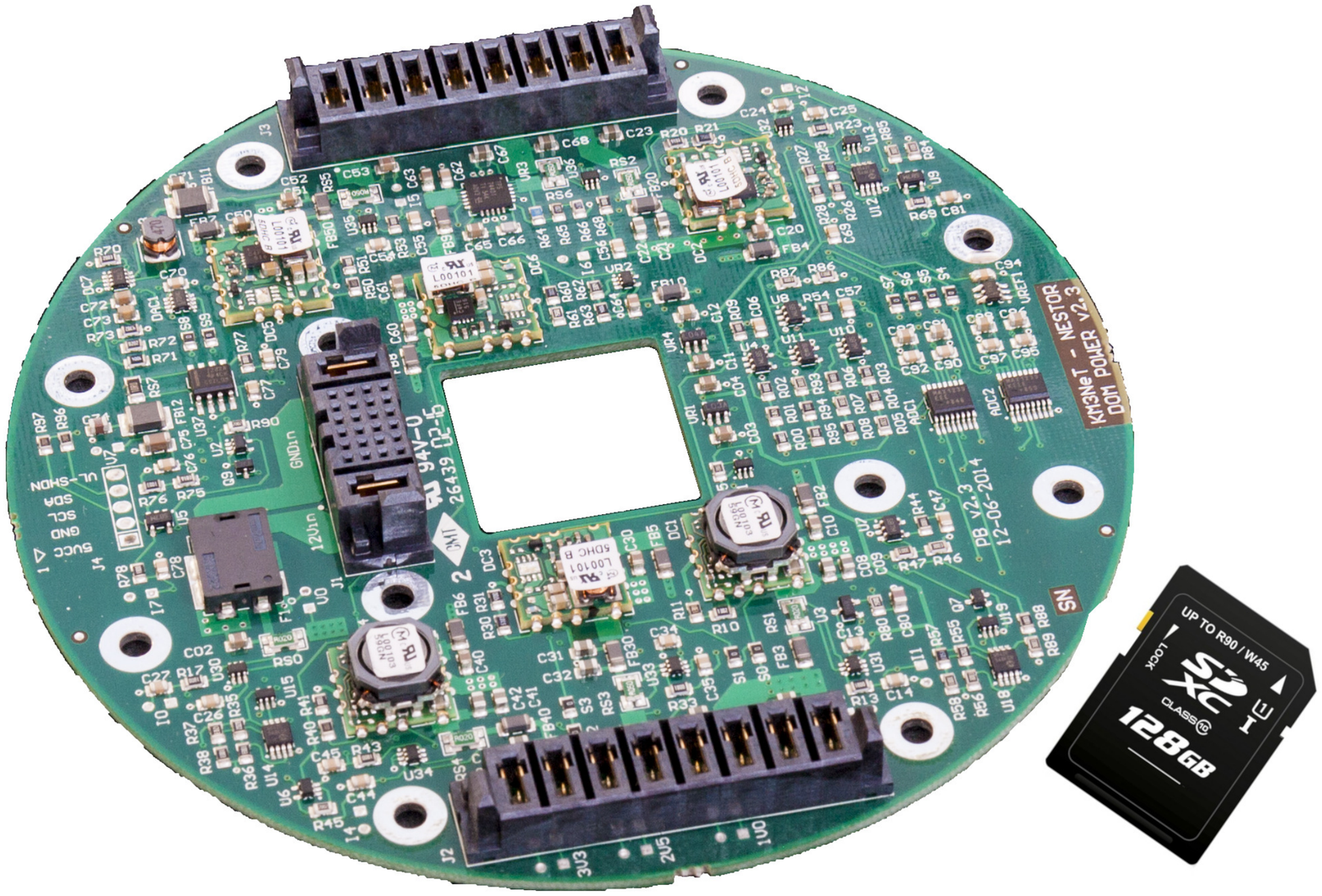}
\caption{The \gls{DOM} Power Board. An SD memory is presented close to the Power Board to provide a size reference.}
\label{fig:pbpro}
\end{figure}

\begin{figure}[tbp] % figures (and tables) should go top or bottom of
                    % the page where they are first cited or in
                    % subsequent pages
\centering
\includegraphics[trim={1cm 38.5cm 0.7cm 3.5cm},clip,width=1.05	\textwidth]{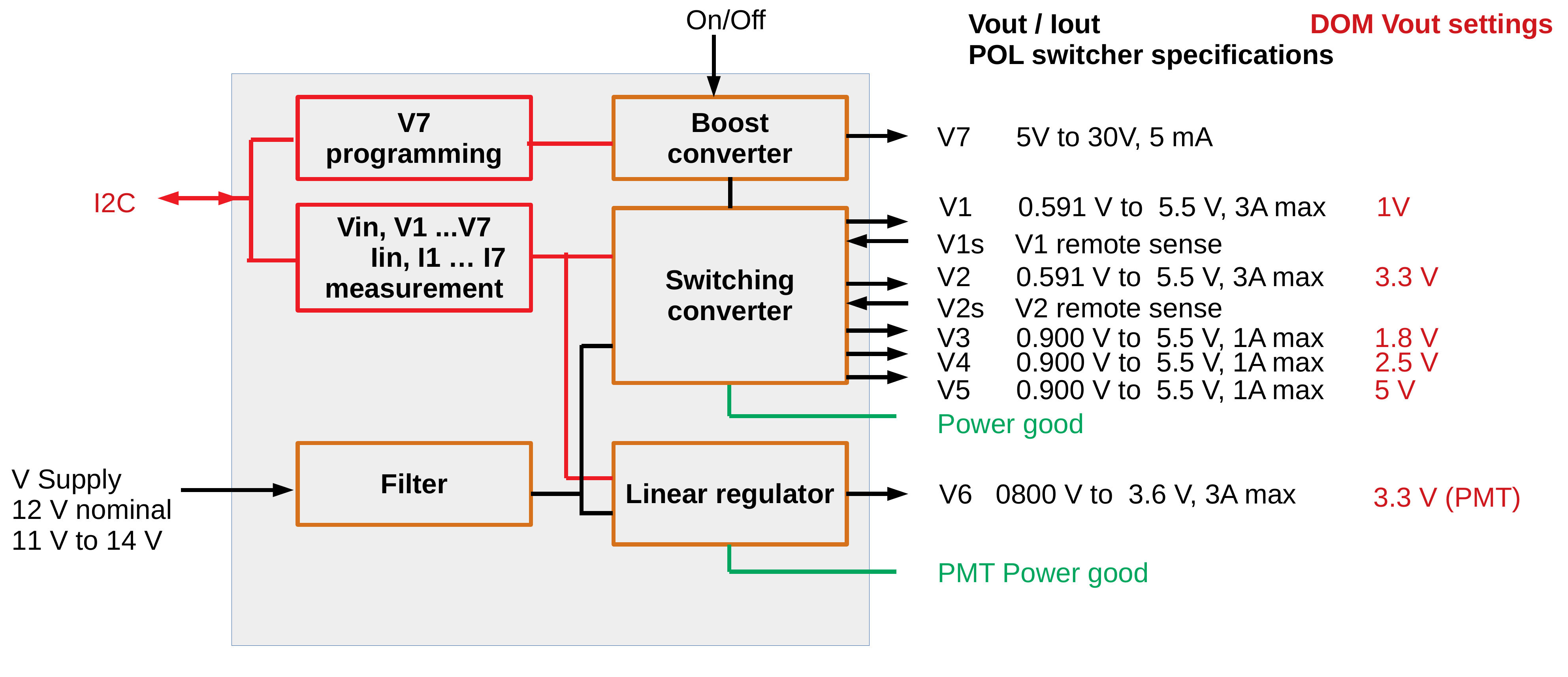}
\caption{Block diagram of the Power Board functionality. The specification of the DC/DC converters are presented for each power rail. The linear regulator included in the Power Board is used to provide an stable voltage to the \glspl{PMT}.}
\label{fig:pbfunc}
\end{figure}

% -------------------------------------------------------------------------------------------
The Power Board,~\cite{pb} shown in  Figure~\ref{fig:pbpro} provides power to the \gls{CLB} and the full \gls{DOM}. The schematic view of the \acrlong{PB} functionality is shown in Figure~\ref{fig:pbfunc}. The input supply to the Power Board is 12~V. Six regulated voltages (1~V, 1.8~V, 2.5~V, 3.3~V, 3.3~V PMT, and 5~V) are generated from the 12~V using DC/DC converters. The 1~V, 1.8~V, 2.5~V and 3.3~V outputs are used by the \gls{CLB} to supply the \gls{FPGA}. The 3.3~V \gls{PMT} output supplies the 31~\gls{PMT} base boards and the 5~V voltage is used to supply the acoustic piezo sensor. Moreover, the \acrlong{PB} provides another output, settable via an \gls{I2C} \gls{DAC}, which results in a configurable voltage ranging from 0~V to 30~V. The settable channel is used by the~Nanobeacon. The Power Board uses high efficiency DC/DC converters in order to minimize the power consumption in the \gls{DOM}. The efficiencies of these DC/DC converters are listed in Table~\ref{tab:tpb}.

\begin{table}[tbp]
\caption{Power Board efficiency for each output voltage.}
\label{tab:tpb}
\smallskip
\centering
\begin{tabular}{ |p{2cm}|p{2cm}|p{3.5cm}|p{2.3cm}|}

 \hline
\centering \textbf{V(V)} &\centering \textbf{I(A)} &\centering  \textbf{Type of DC/DC} &\multicolumn{1}{|c|}{\textbf{Efficiency (\%)}}
\\
\hline
\centering 2.5 &\centering 0.13&\centering LTM8021& \multicolumn{1}{|c|}{80}
\\
\hline
\centering 3.3 &\centering 0.33&\centering OKL-1& \multicolumn{1}{|c|}{90}  \\
 \hline
\centering 3.3 & \centering 0.34&\centering OKL-1& \multicolumn{1}{|c|}{90}  \\
 \hline
\centering 1.0 &\centering 0.80&\centering OKL-3& \multicolumn{1}{|c|}{80}  \\
 \hline
\centering 1.8 &\centering 0.46&\centering OKL-3& \multicolumn{1}{|c|}{80}  \\
\hline
\centering 5.0 &\centering 0.10&\centering MAX17542G& \multicolumn{1}{|c|}{90}  \\

 \hline
\end{tabular}
\end{table}

In order to protect the sensitive electronics inside the \gls{DOM} from the interferences by the high frequency noise produced by the DC/DC converters, the Power Board is located in the shielded part of the cooling mushroom. The chosen location provides also a better cooling of the Power Board. The location of the Power Board in the \gls{DOM} is shown in Figure~\ref{fig:dom2d}.


\begin{figure}[tbp] % figures (and tables) should go top or bottom of
                    % the page where they are first cited or in
                    % subsequent pages
\centering
\includegraphics[trim={2cm 1cm 0cm 0cm},clip,width=1.15\textwidth]{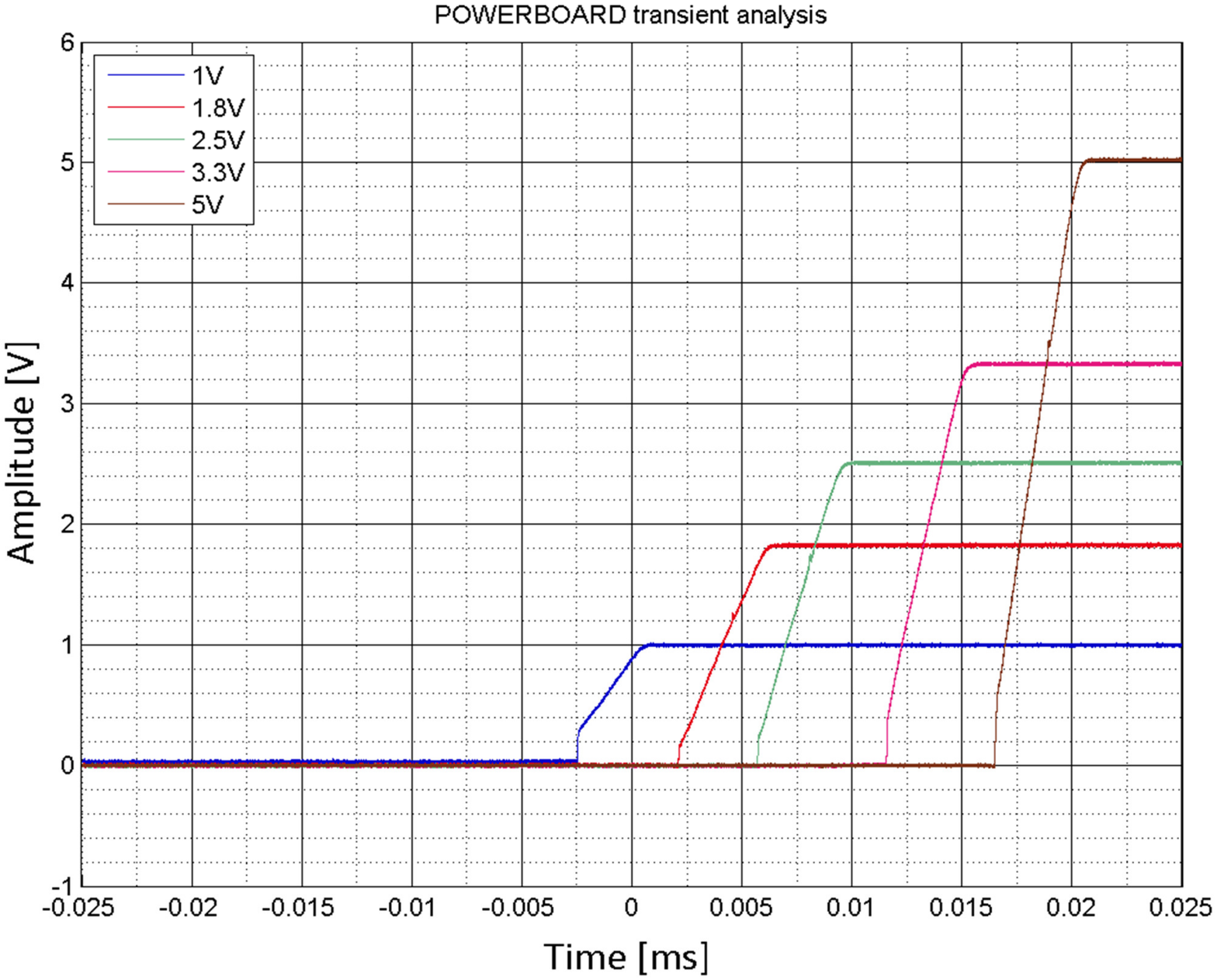}
\caption{Power up sequence. The picture corresponds to an oscilloscope capture. The oscilloscope trigger, set to 0.9 V, fixes the time reference.}
\label{fig:powerup}
\end{figure}

% -------------------------------------------------------------------------------------------
\subsection{Power startup}\label{sec:pwrstartup}
% -------------------------------------------------------------------------------------------
One of the functions of the Power Board is to provide  a proper voltage startup sequence to the \gls{FPGA}. For this purpose, a sequencer has been implemented in the Power Board in order to provide the needed sequence of voltages\cite{powerup}. The sequence of voltages generated by the Power Board is shown in Figure~\ref{fig:powerup}. Two power-good signals are generated by the Power Board. The first one indicates that the 3.3~V~\gls{PMT} output has been successfully started (power-good \gls{PMT}). The second one indicates the successful completion of the power up sequence. The final function implemented in the Power Board is a hysteresis loop to avoid instabilities at the startup. The regulators of the Power Board are enabled only when the input voltage exceeds 11~V, whereas they are disabled when the input value drops below 9.5 V. In this way, fluctuations in the Power Board regulators are avoided at the start point of the input voltage.

\section{Photomultiplier Base}\label{sec:pmtbase}
\begin{figure}[tbp]
\includegraphics[trim={0cm 0cm 0cm 0cm},clip,angle=90,width=1\linewidth]{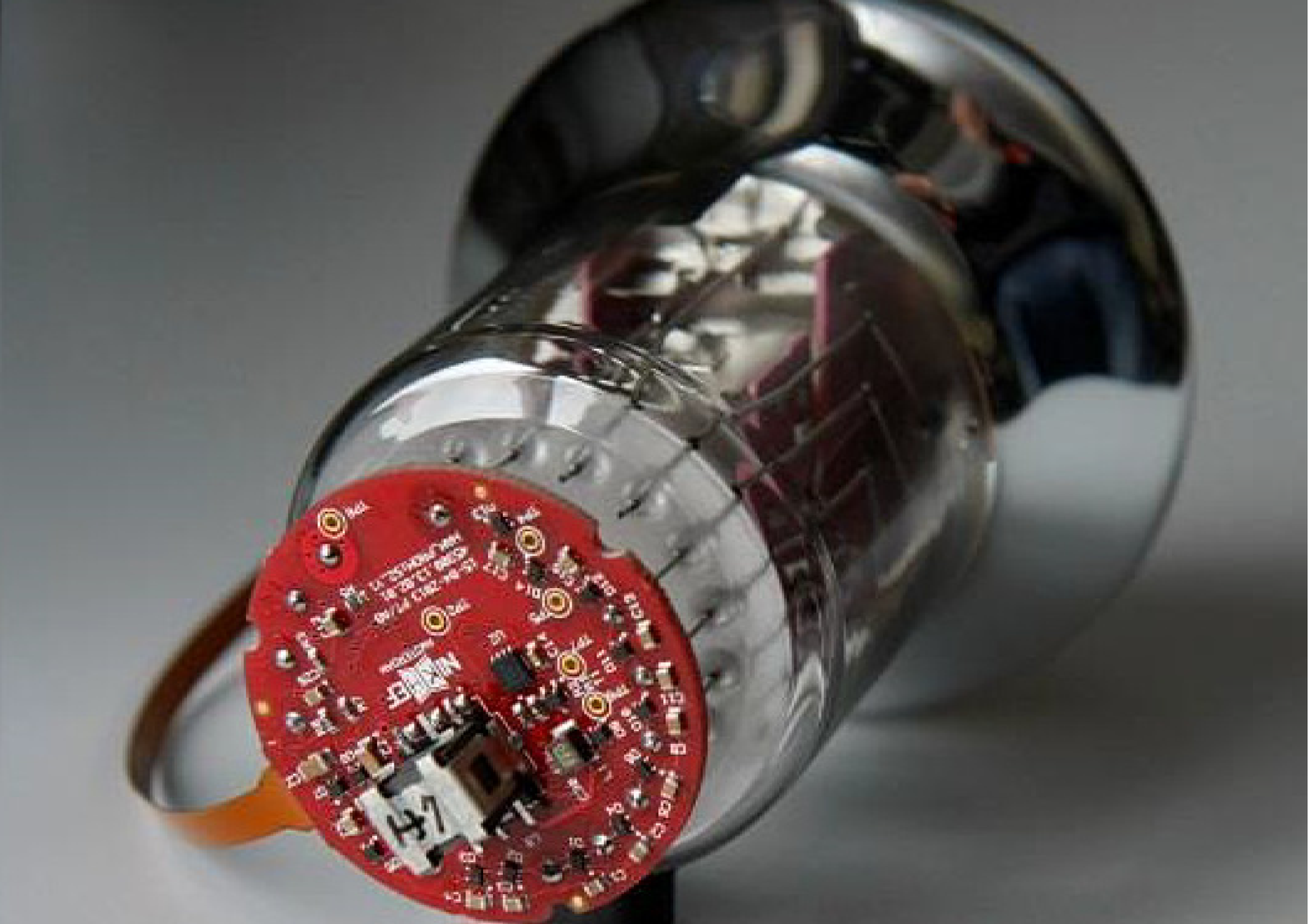}
\caption{The \gls{PMT} base board mounted on a 3-inch \gls{PMT}.}
\label{fig:subpmtboard}
\end{figure}

\begin{figure} [tbp]
\includegraphics[trim={0cm 0cm 0cm 0cm},clip,width=1\linewidth]{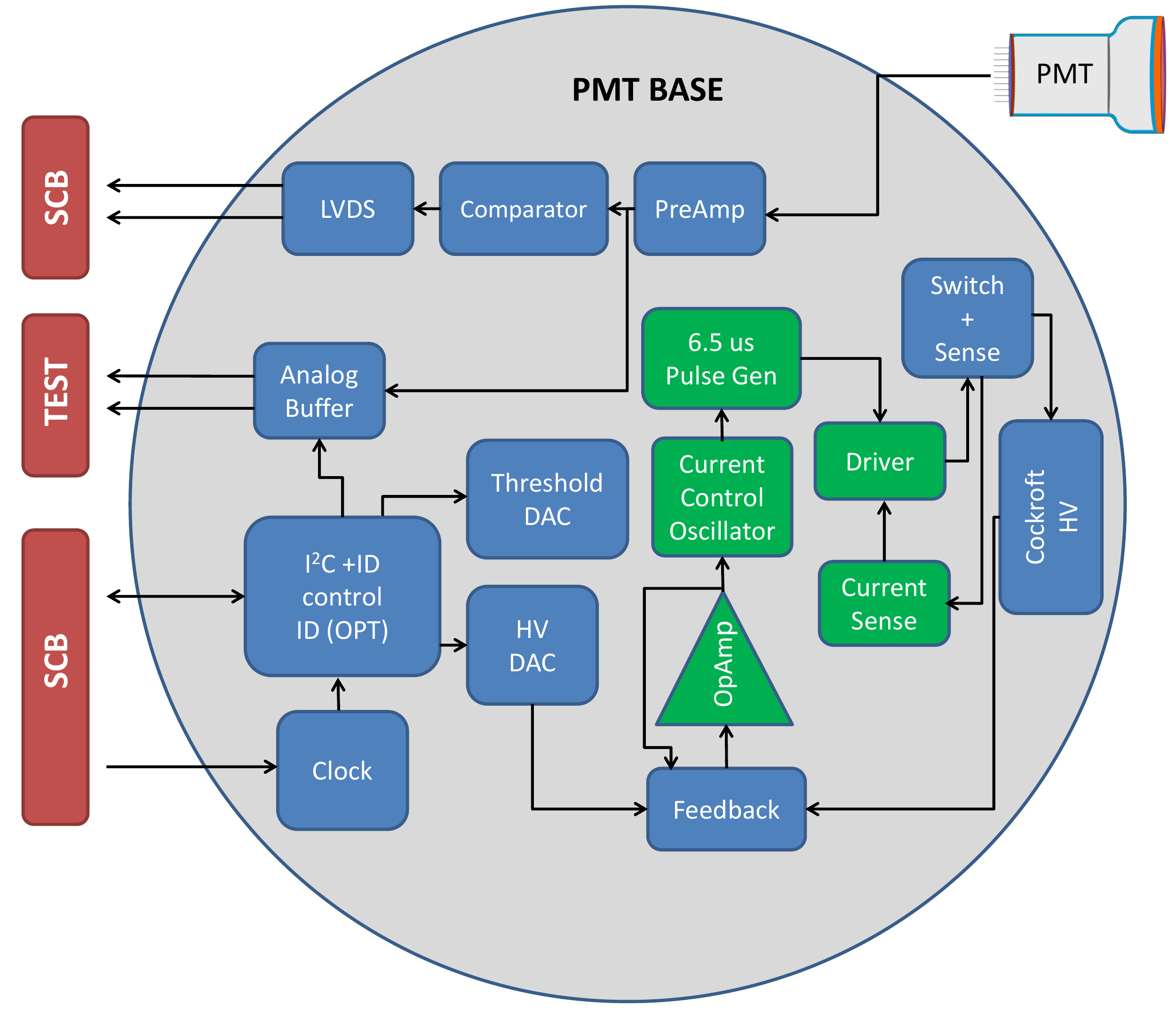}
\caption{Block diagram of the \gls{PMT} base.}
\label{fig:subpmtdiag}
\end{figure}

% -------------------------------------------------------------------------------------------
The \gls{PMT} base board~\cite{pmtbase} (see Figure~\ref{fig:subpmtboard}) takes care of both the generation of the \gls{HV} supplied to the \gls{PMT} and the digitization of the \gls{PMT} signal. Before being digitized, the \gls{PMT} signal is amplified by a pre-amplifier built in the \gls{PMT} base. One of the main components of the \gls{PMT} base is a comparator, which provides a logical high signal when the \gls{PMT} output is over the comparator threshold -set through \gls{I2C}-. The duration of the primary signal (\gls{ToT}) provided by the \gls{PMT} bases is accurately measured by the \gls{CLB} \glspl{TDC}. Apart from the logical signal, the \gls{PMT} base outputs also the amplified analogue \gls{PMT} signal, which is only used for testing. The 31 \gls{PMT} base boards are connected to the \gls{SCB} by flexible \gls{PCB}. The \gls{HV}, which is remotely configurable through \gls{I2C}, is independently generated in each \gls{PMT} base. This allows for tuning the gain of individual \gls{PMT}s in order to equalize cross-\gls{PMT} photon response. The \gls{HV} value can be adjusted remotely, from $-$800 to $-$1400~V. Figure~\ref{fig:subpmtdiag} shows a diagram of the PMT base board with its main components.

% -------------------------------------------------------------------------------------------
\subsection{Photomultiplier Base \acrshort{ASIC}s}\label{sec:pmtbaseasic}
% -------------------------------------------------------------------------------------------

In order to reduce the space occupied by the \gls{PMT} base, as well as its cost and power consumption, two \glspl{ASIC} have been developed~\cite{asic}. As the \gls{DOM} is tightly packed with the 31 \gls{PMT}s and the electronics, compactification is crucial. The first \gls{ASIC} is the so-called PROMiS \gls{ASIC}, which performs the readout of the \gls{PMT} signals and has two different parts, one digital and one analog. The second chip is the CoCo \gls{ASIC}, which controls the Cockroft-Walton \gls{HV} power supply providing a gain of $10^6$. The main characteristics of both \gls{ASIC}s are listed in Tables \ref{tab:tbas1} and \ref{tab:tbas2}.

\begin{table}[tbp]
\caption{Specifications of the PROMiS chip.}
\label{tab:tbas1}
\smallskip
\centering
\begin{tabular}{ |p{8cm}|p{5cm}| }

 \hline
Time resolution (for a single photon, photomultiplier + electronics) &  \multicolumn{1}{|c|}{$<$ 2ns} \\
  \hline
Two-hit time separation &  \multicolumn{1}{|c|}{$\geq$ 25 ns}  \\
 \hline
 Power consumption  &  \multicolumn{1}{|c|}{    35 mW}\\
 \hline
 Supply voltage, Technology  &  \multicolumn{1}{|c|}{ 3.3 V, 0.35 $\mu$m CMOS AMS} \\
 \hline
Comparator Threshold Adjustment  &  \multicolumn{1}{|c|}{8 bits (0.8 V - 2.4 V) }\\
 \hline
\gls{HV} feedback control   & \multicolumn{1}{|c|}{    8 bits (0.8 V - 2.4 V)}\\
\hline
Slow-Control Communication, Digital and Analog Output   &  \multicolumn{1}{|c|}{   \gls{I2C}, \gls{LVDS} and Analog buffer respectively } \\
 \hline
\end{tabular}
\end{table}

\begin{table}[tbp]
\caption{Specifications of the CoCo chip.}
\label{tab:tbas2}
\smallskip
\centering
\begin{tabular}{ |p{8cm}|p{5cm}| }
 \hline
Pulse output frequency &\multicolumn{1}{|c|}{ $<$ 50 kHz (max.)} \\
  \hline
Pulse width & \multicolumn{1}{|c|}{$<$ 6.5 $\mu$s (max.)}  \\
 \hline
 Power consumption  &  \multicolumn{1}{|c|}{ $ <$ 1 mW}\\
 \hline
 Supply voltage, Technology  &\multicolumn{1}{|c|}{3.3 V, 0.35 $\mu$m CMOS AMS }\\
 \hline
Current sense & \multicolumn{1}{|c|}{100 mV over 1.5 $\Omega$ } \\
 \hline
Operational amplifier reference (internal)  &  \multicolumn{1}{|c|}{1.2 V} \\

 \hline

\end{tabular}
\end{table}

\begin{figure}[tbp] % figures (and tables) should go top or bottom of
                    % the page where they are first cited or in
                    % subsequent pages
\centering
\includegraphics[trim={7.15cm 6cm 9.5cm 2cm},clip,width=1\textwidth]{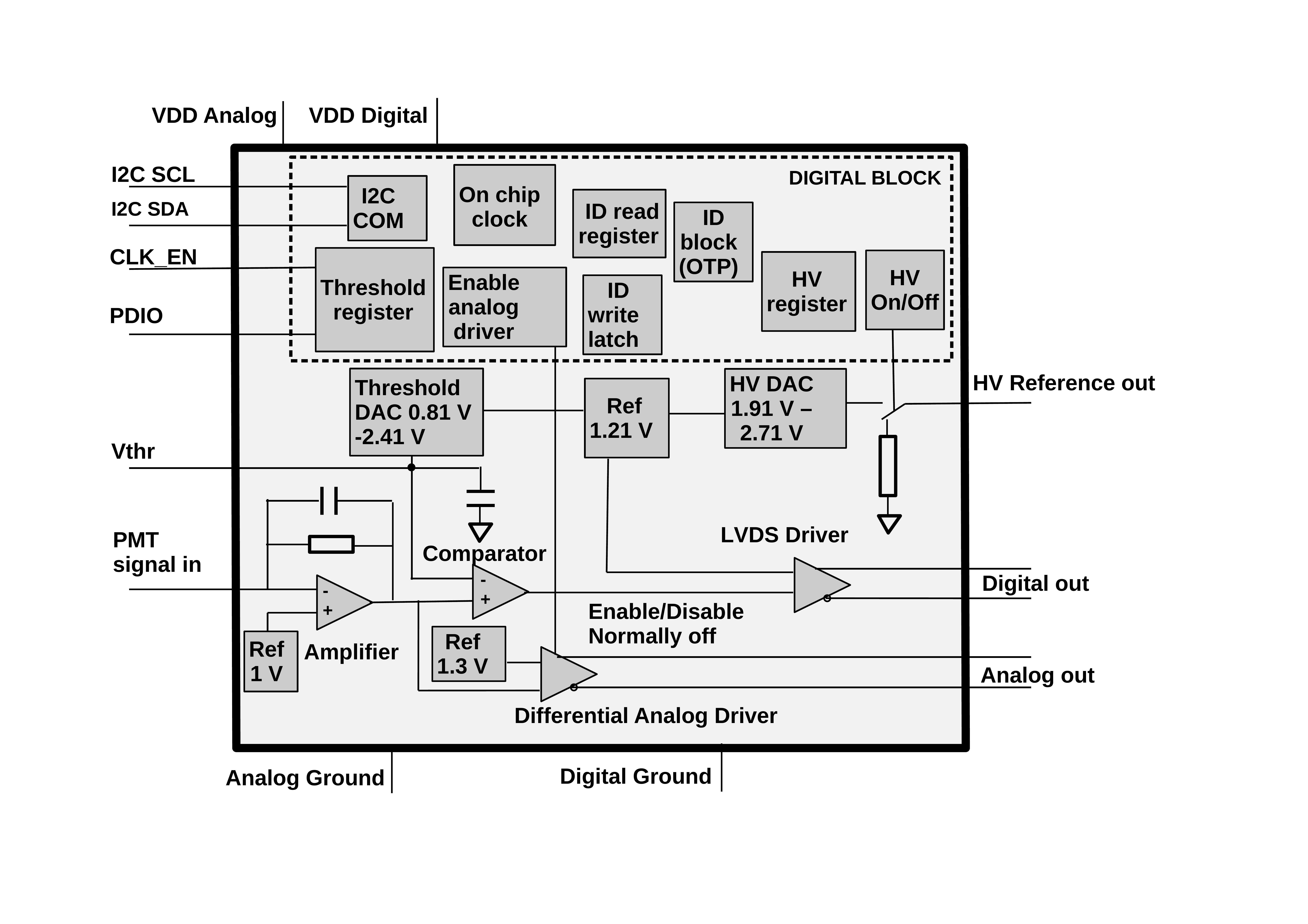}
\caption{Diagram of the PROMiS chip.}
\label{fig:dpro}
\end{figure}

% -------------------------------------------------------------------------------------------
\subsubsection{PROMiS Analog block}\label{sec:analogprms}
% -------------------------------------------------------------------------------------------
The analog section of the PROMiS \gls{ASIC} includes: a pre-amplifier, which can increase the amplitude signal of the \gls{PMT};  a two-stage charge amplifier biased at 1~V (feedback: R\textsubscript{f}~=~15~k$\Omega$,   C\textsubscript{f} = 300 fF); and a discriminator which compares the input signal with a predefined threshold level configured by \gls{I2C}. The PROMiS \gls{ASIC} generates the \gls{LVDS} signal. The signal is transmitted via the \gls{SCB} to the \gls{CLB} where it is digitized by the corresponding \gls{TDC}. The \gls{LVDS} driver, with common mode feedback, feeds the 100~$\Omega$ kapton transmission line. The 1.2~V reference voltage is produced by the band gap. From this reference voltage, all the remaining voltages and currents are generated. Figure~\ref{fig:dpro} shows the block diagram of the \gls{ASIC}. 

% -------------------------------------------------------------------------------------------
\subsubsection{PROMiS Digital block}\label{sec:digitprms}
% -------------------------------------------------------------------------------------------
The digital block of the PROMiS \gls{ASIC} includes a clock generator that produces a 10~MHz clock signal with possible fluctuations due to temperature and voltage up to 30~\%. The clock accuracy is not critical as it is used by \gls{I2C} interface, whose bus operates at 250~kHz. The PROMiS \gls{ASIC} also includes an \gls{I2C} slave and  \gls{OTP} memory block. In Figure~\ref{fig:dpro} the block diagram of the PROMiS \gls{ASIC}, including its digital part, is shown. In order to save power, it is possible to shut off the clock via an enable/disable signal. The \gls{ASIC} also provides the possibility to test the analog chain via \gls{I2C} as well as to switch on and off the \gls{HV} circuit. 

\begin{figure}[tbp] % figures (and tables) should go top or bottom of
                    % the page where they are first cited or in
                    % subsequent pages
\centering

\includegraphics[trim={4.5cm 8.5cm 1cm 0.5cm},clip,width=1\textwidth]{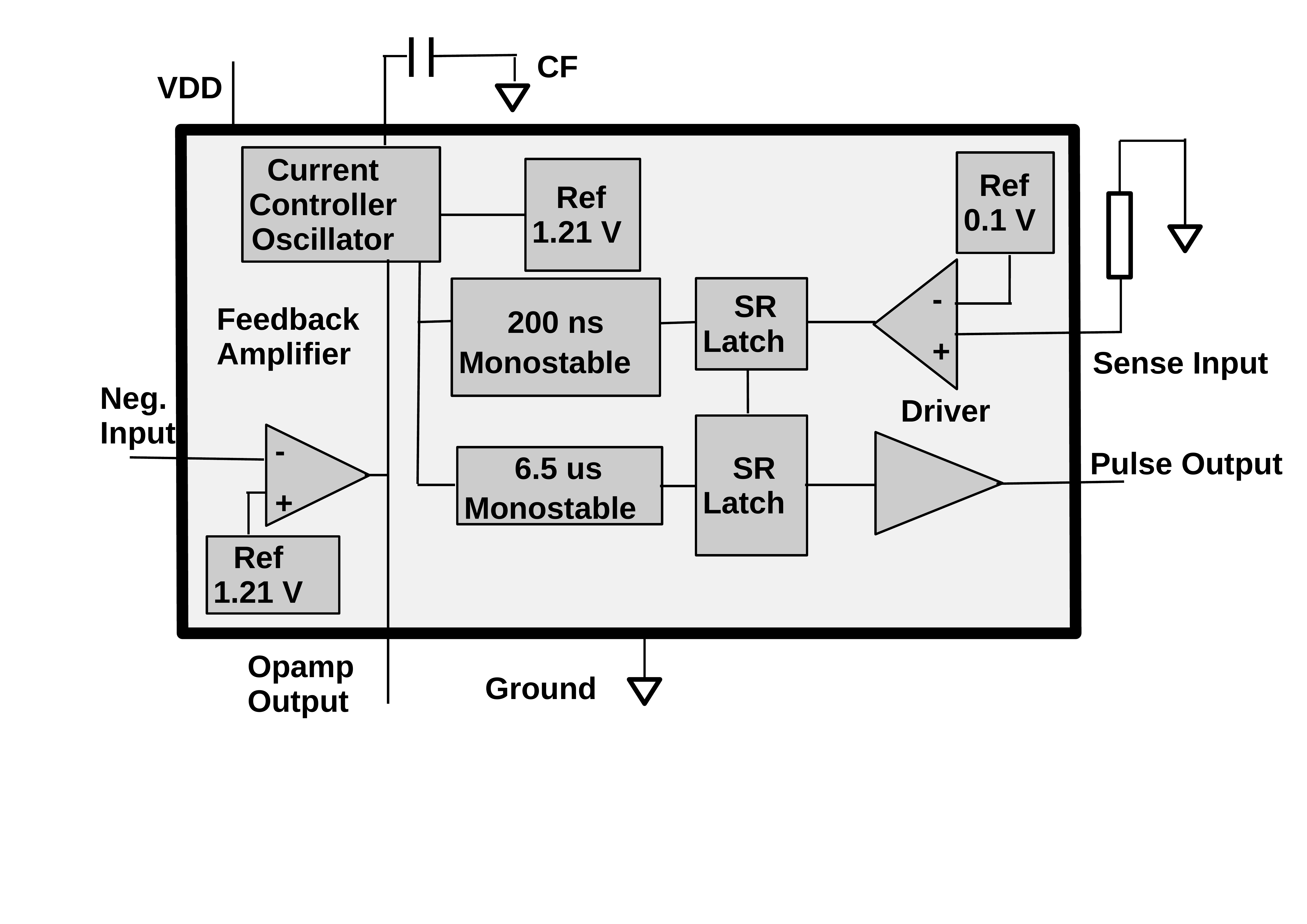}
\caption{Block Diagram of the CoCo chip.}
\label{fig:coco}
\end{figure}

% -------------------------------------------------------------------------------------------
\subsubsection{CoCo - Cockroft Walton multiplier feedback control \gls{ASIC}}\label{sec:cocofb}
% -------------------------------------------------------------------------------------------
The CoCo \gls{ASIC} (see block diagram in Figure~\ref{fig:coco}) controls the autotransformer of the \gls{PMT} base. The autotransformer, which has a ratio of 1:12, couples the 3.3~V power supply provided by the \gls{SCB} to the \gls{CW} multiplier circuit. The \gls{CW} multiplier circuit generates the stable \gls{HV} needed by the \gls{PMT}, whose gain has a linear response to the \gls{HV}. The \gls{ASIC} receives feedback from the \gls{CW} multiplier circuit in order to accurately control  the \gls{HV}. The control is performed by a series of pulses to the switch that is managing the autotransformer. The characteristic pulse width is 6.5~$\mu$s and its frequency, which determines the \gls{HV}, changes according to the \gls{HV} feedback.  The \gls{HV} feedback voltage is used for charging (or discharging) a capacitor. The value of the capacitor, loaded by the current of the \gls{HV} feedback, sets the frequency. The triangular wave created by the charge and discharge of the capacitor is also used for generating internal clocks. Another function of the \gls{HV} feedback is to avoid the autotransformer saturation in case of short circuit.

\begin{figure}[tbp] % figures (and tables) should go top or bottom of
                    % the page where they are first cited or in
                    % subsequent pages
\centering
\includegraphics[width=\textwidth]{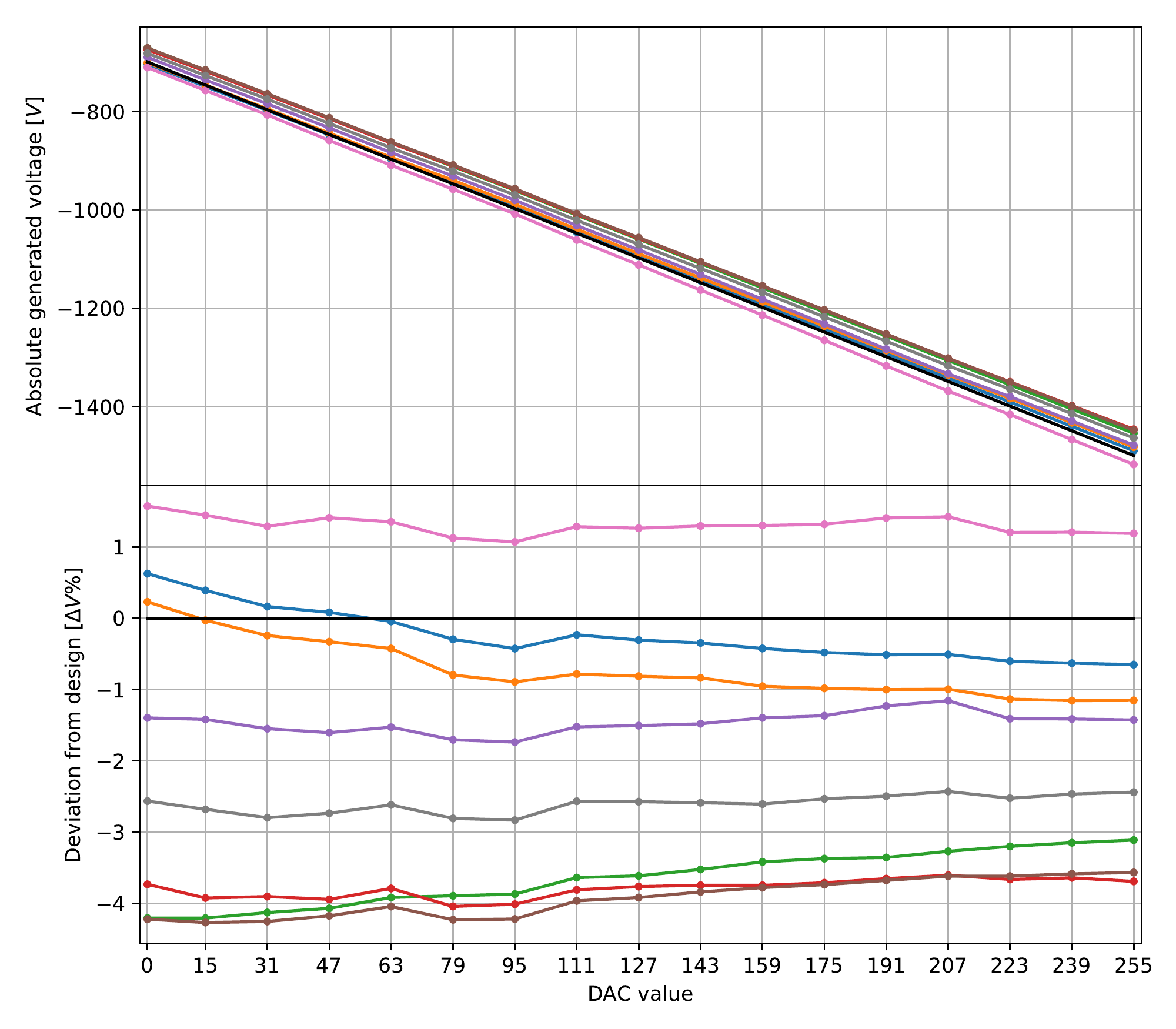}
\caption{\gls{HV} produced by a sample of \gls{PMT} Bases against set PROMiS \gls{DAC} value. The top plot shows the absolute voltage for eight selected bases, while bottom plot shows the relative deviation from design voltage, where ${\Delta}V\%= {}^{(V_{DAC}-V_{\text{design}}})/{}_{V_{DAC}} \times 100$.}
\label{fig:pmthv}
\end{figure}
The relation between the \gls{PMT} \gls{DAC} value and the actual produced \gls{HV} is shown in Figure~\ref{fig:pmthv}. The relation between the \gls{DAC} value and the output voltage can be derived from the \gls{PMT} base \gls{HV} circuit and is given by
\begin{equation}
V_{HV} = -F(V_{min} - V_{ref} + D\frac{V_{max}-V_{min}}{255}) + V_{ref} 
\end{equation}
where $D$ is the \gls{DAC} value (0-255), $V_{ref}$ is the reference voltage generated by the Cockroft Walton multiplier feedback control \gls{ASIC} (1.21~V), $V_{min}$ and $V_{max}$ are the minimum and maximum output voltages of the \gls{DAC} (1.91~V and 2.71~V respectively) and $F$ is the feedback path voltage divider factor, which has been set to 1000. $V_{HV}$ is the \gls{HV} generated by the circuit. From this, it follows a range variation of the design output between $-$698.8~V and $-$1498.8~V for \gls{DAC} values 0 and 255 respectively. The observed variation among a sample of \gls{PMT} bases in Figure~\ref{fig:pmthv} is due to resistor tolerances present in the feedback loop, which total to a maximum of $\pm 6$\%. The nonlinearity and general offset with respect to the design voltage in the plot is due to the inherent error of the measurement method.

% -------------------------------------------------------------------------------------------
\section{The \acrlong{SCB}}\label{sec:oct}
\begin{figure}[tbp] % figures (and tables) should go top or bottom of
                    % the page where they are first cited or in
                    % subsequent pages
\centering

\begin{subfigure}[b]{.49\linewidth}
\includegraphics[trim={4cm 5cm 2cm 13.5cm},clip,width=1\linewidth]{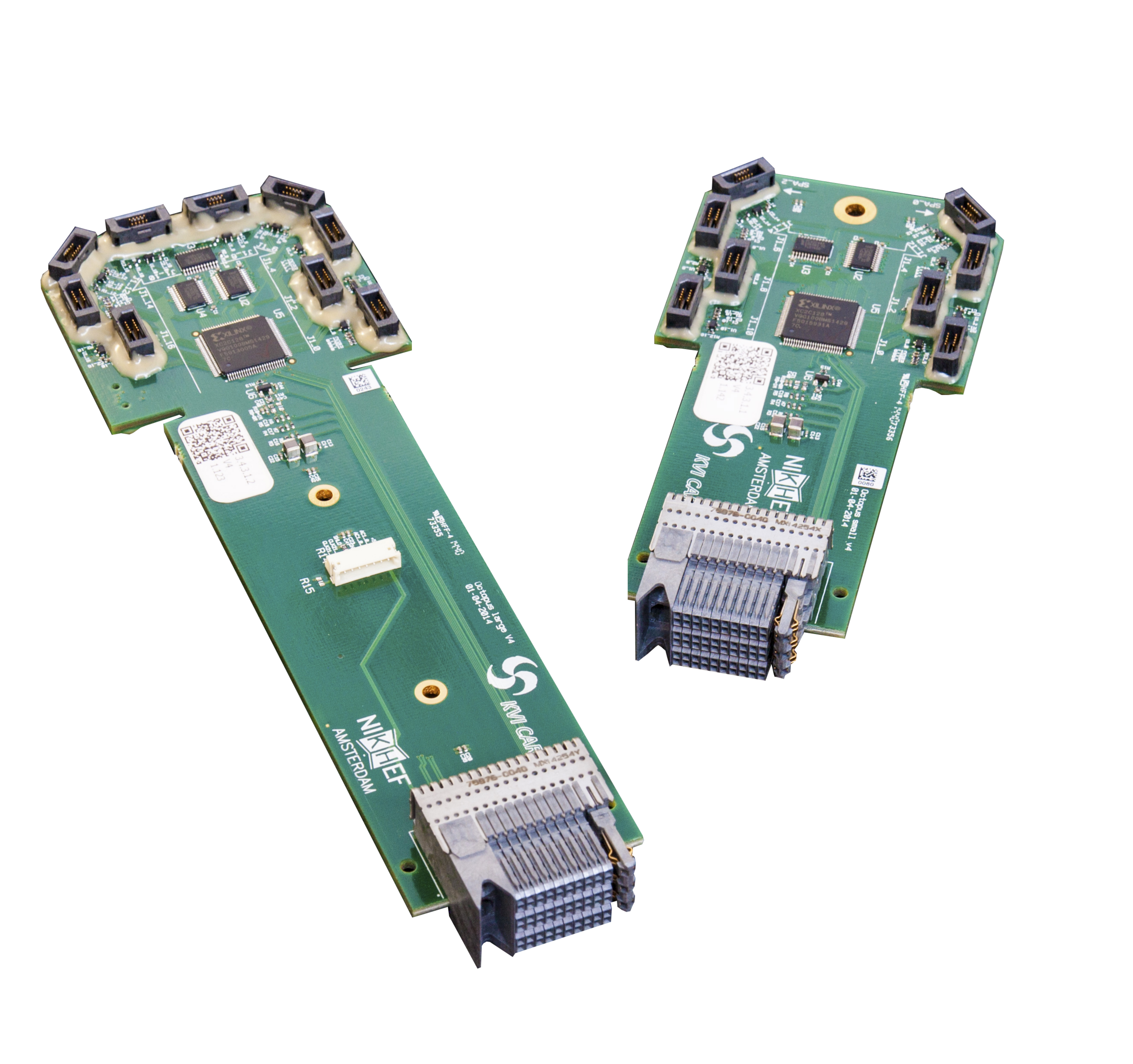}
\caption{The two \gls{SCB} of the \gls{DOM}.}
\label{fig:octa}
\end{subfigure}
\begin{subfigure}[b]{.49\linewidth}
\includegraphics[trim={34cm 44cm 46cm 22cm},clip,width=1\linewidth]{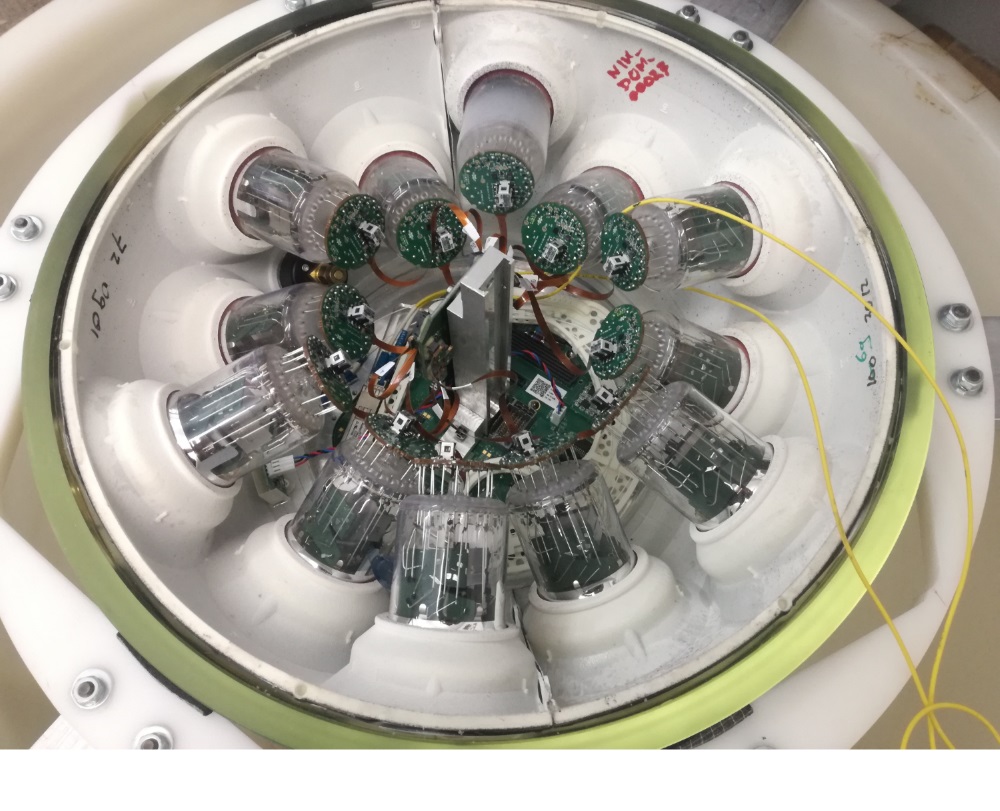}
\caption{An \gls{SCB} inserted in the \gls{DOM}.}
\label{fig:octb}
\end{subfigure}

\caption{Signal Collector Boards (\gls{SCB}).}
\label{fig:oct}
\end{figure}

% -------------------------------------------------------------------------------------------
The \gls{PMT} base generated \gls{LVDS} signals are collected on a hub board, called the \gls{SCB}. The main function of the \gls{SCB} is to transfer the signals from the \gls{PMT} base to the \gls{TDC}s embedded in the \gls{CLB}. The \gls{SCB} also transfers the \gls{I2C} command signals from the \gls{CLB} to the \gls{PMT} bases, in order to monitor and control the \glspl{PMT}. Each \gls{DOM} comprises two \glspl{SCB}, one large and one small (Figure~\ref{fig:octa}). Figure~\ref{fig:octb} shows one \gls{SCB} connected to the \gls{PMT} bases in half a \gls{DOM}. 
The architecture of the \gls{SCB} consists of the following parts:
\begin{itemize}
\item Backplane connector to the \gls{CLB}.
\item Xilinx Coolrunner \gls{CPLD}.
\item \gls{I2C} multiplexer.
\item Current limit switches.
\item \gls{PMT} channels: 19 in the large \gls{SCB} and 12 in the small \gls{SCB}.
\item One piezo connector  (only in the large \gls{SCB}).
\end{itemize}
\gls{LVDS} signaling, as used between the \gls{PMT} base and the \gls{CLB}, is not susceptible to cross-talk due to the fact that the two signal lines of the \gls{LVDS} are electrically tightly coupled with matched impedance throughout the complete route from the \gls{PMT} base to the \gls{CLB}. The signal that can be coupled into the \gls{LVDS} line will be coupled into both signal lines at the same time. Because of this, the distortion becomes common mode and will not affect the signal integrity. For each \gls{PMT}, a re-settable fuse \gls{IC}, integrated on the \gls{SCB}, protects individual \gls{PMT}s and the \gls{CLB} from short circuit or excesive current draw. For control and monitoring of the \gls{SCB}, a \gls{CPLD}, accessible through \gls{I2C}, has been added. The \gls{CPLD} allows for reading and resetting the current sensors and disabling the \gls{PMT} base digital clock to eliminate possible interferences from this clock on the \gls{PMT} signals. The acoustic piezo sensor is also connected to the \gls{CLB} via the large \gls{SCB}. As in the case of the \gls{PMT}s, the \gls{SCB} supplies the piezo with the needed voltage and transfers the acquired data from the piezo sensor to the \gls{CLB}. The piezo does not feature a control interface.

% -------------------------------------------------------------------------------------------
%\subsection{\gls{SCB} Architecture}\label{sec:archdecomp}
% -------------------------------------------------------------------------------------------

\begin{figure}[tbp] % figures (and tables) should go top or bottom of
                    % the page where they are first cited or in
                    % subsequent pages
\centering

\includegraphics[trim={0.5cm 1cm 2cm 2cm},clip,width=1.05\textwidth]{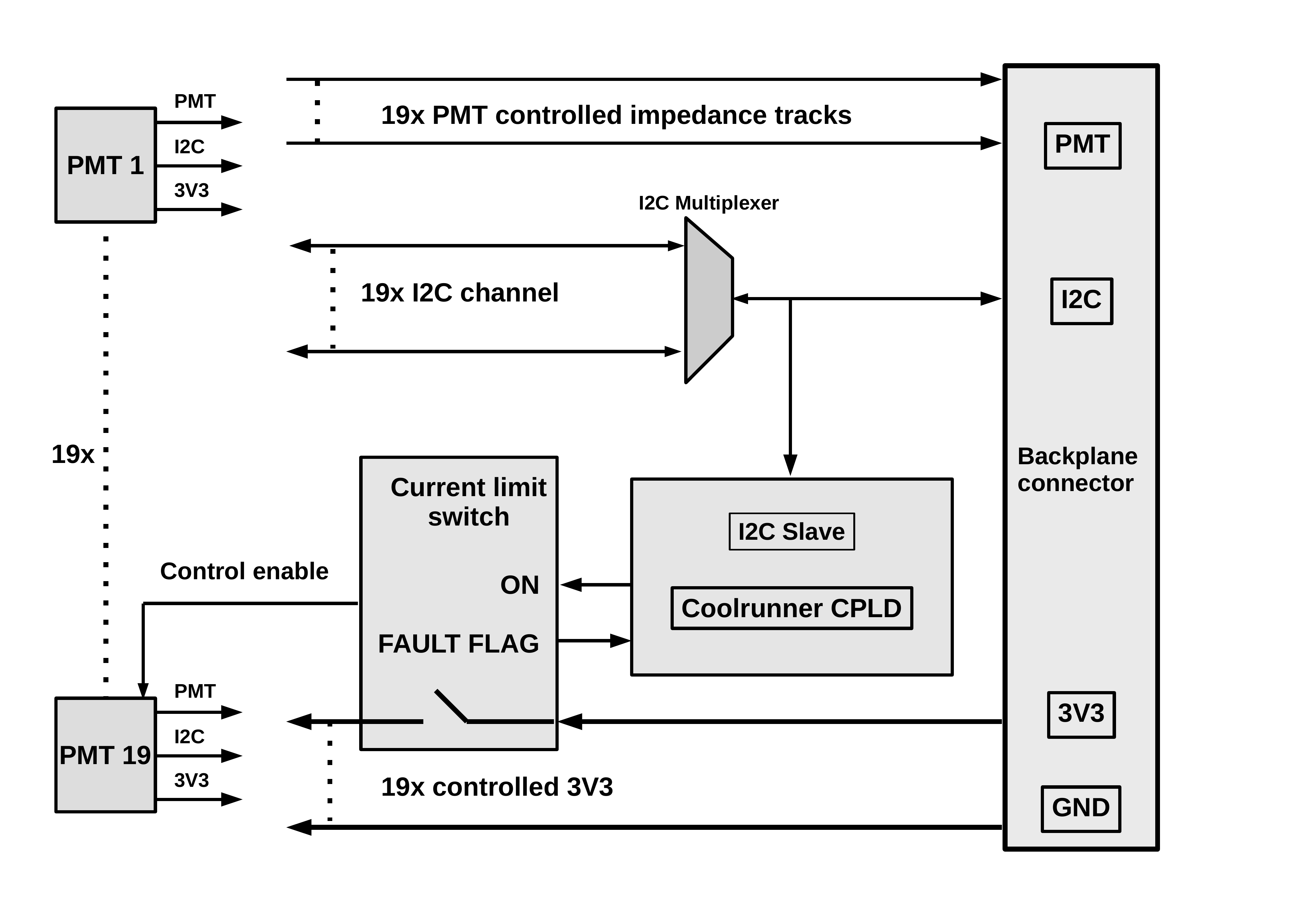}
\caption{Block diagram of the 19 \gls{PMT} channel interface of the large \gls{SCB}. The small \gls{SCB} is analogous with only 12 channels.}
\label{fig:octhard}
\end{figure}

The large \gls{SCB} has 19~equal channels (see Figure~\ref{fig:octhard}). The \gls{LVDS} signals and the~5~V needed to supply  the acoustic piezo sensor are connected from the backplane connector to the piezo connector. 

The~5~V power is not measured and cannot be switched by the \gls{SCB}. 
The small \gls{SCB} has~12~\gls{PMT} channels and three spare channels.

% -------------------------------------------------------------------------------------------

% -------------------------------------------------------------------------------------------
\section{\gls{DOM} Power budget}\label{sec:con}
% -------------------------------------------------------------------------------------------
The power consumption breakdown of the most consuming \gls{DOM} electronics boards is shown in Table~\ref{tab:tcon}. The component of the \gls{DOM} with the highest power consumption is the \gls{CLB}. Inside the \gls{CLB}, the \gls{FPGA} and the \gls{SFP} are the main power consumers, followed by the clock conditioner and the Nanobeacon. The Nanobeacon is only operated when a calibration run is performed, typically a few minutes once a week.

\begin{table}[tbp]
\caption{\gls{DOM} power budget.}
\label{tab:tcon}
\smallskip
\centering
\begin{tabular}{ |p{5cm} | p{5cm} | p{2cm}| }
 \hline
 \multicolumn{3}{|c|}{Power budget} \\
 \hline
Board & Sub-component & Power (W)\\
 \hline
 Power Board & &  \multicolumn{1}{|c|}{0.72} \\
 Central Logic Board   &      & \multicolumn{1}{|c|}{4.45}\\
               & \textit{\gls{FPGA}}  &  \multicolumn{1}{|l|}{\textit{2.25}}\\
               &\textit{\gls{SFP}} &  \multicolumn{1}{|l|}{\textit{1.50}}\\
               &\textit{Clock conditioner} & \multicolumn{1}{|l|}{\textit{0.50}} \\
               
              % & Temp + humidity & 0.015   \\
               &\textit{Tilt and compass}  & \multicolumn{1}{|l|}{ \textit{0.20}}\\
Small \gls{SCB}  & & \multicolumn{1}{|c|}{0.02} \\
         %     &  OM \gls{I2C} circuit  & 0.02\\
            %   & OM \gls{I2C} circuit & 0.00078\\
Large \gls{SCB} & & \multicolumn{1}{|c|}{0.02}\\
           %    &  OM \gls{I2C} circuit  & 0.02\\
             %  & OM \gls{I2C} circuit & 0.00078\\
\gls{PMT} 31$\times$ & & \multicolumn{1}{|c|}{1.05} \\
            %   &  HV  & 0.004\\
            %   & Detection & 0.03\\
Digital Piezo & & \multicolumn{1}{|c|}{0.50}\\
\hline
Total Budget & & \multicolumn{1}{|c|}{6.76}\\

 \hline
\end{tabular}
\end{table}

The \gls{SCB} consumption is negligible and the 31 \gls{PMT} bases add up to a total of 1~W. The Power Board, mainly because of the DC/DC converter losses, accounts for 10.2\% of the total \gls{DOM} power consumption. In total, the power consumption of the \gls{DOM} is around 7~W when fully operational. Keeping low the power drain is important both for keeping the overall consumption of the detector low and minimizing the heat production.

% -------------------------------------------------------------------------------------------
\section{Reliability}\label{sec:fides}
% -------------------------------------------------------------------------------------------

Maintenance of \gls{DU} operated in deep seawater is difficult. In order to quantify the reliability of the electronics boards used in the detector, the FIDES\cite{fidesg} method is used. The FIDES methodology provides two main engineering tools. The first one consists of a handbook for predicting the reliability of the electronic boards analysed. The second one is a guide to estimate the impact of the design and manufacturing processes on the reliability of the produced boards. FIDES provides a spreadsheet tool to calculate the \gls{FIT} and the \gls{MTTF} of an electronics board. Given a board, each of its components is assigned a \gls{FIT}, which is either provided by the manufacturer or obtained from the FIDES handbook. The final \gls{FIT} of a board is the sum of the \glspl{FIT} of each single component and estimates the failure rate per $10^9$  hours. Once the \gls{FIT} is obtained, it is possible to calculate the probability of failure in a given time as $F(t) = 1 - R(t)$, being  $R(t)$\ the probability of a system to be still operational over a time period \textit{t}. $R(t)= e^{-\lambda t}$, $\lambda$\ being the board \gls{FIT} value and \textit{t} the time period duration in hours.

\begin{table}[tbp]
\caption{\gls{FIT} and \gls{MTTF} of the \gls{DOM} electronics boards of KM3NeT.}
\label{tab:tabfit}
\smallskip
\centering
\begin{tabular}{ |p{4cm}|p{3cm}|p{3cm}|p{3cm}|}
\hline
Product  &  \multicolumn{1}{|l|}{FIT} & \multicolumn{1}{|l|}{MTTF(years)}\\
\hline
\gls{PMT} Base
& \multicolumn{1}{|c|}{1218}  & \multicolumn{1}{|c|}{94}\\
\hline
Large \gls{SCB}& 
 \multicolumn{1}{|c|}{157}   &   \multicolumn{1}{|c|}{727}\\
\hline

Small \gls{SCB}&
 \multicolumn{1}{|c|}{156}   & \multicolumn{1}{|c|}{731}\\
\hline
Power Board& 
 \multicolumn{1}{|c|}{1424}   &  \multicolumn{1}{|c|}{80}\\
\hline
\gls{CLB}& 
 \multicolumn{1}{|c|}{417}   &  \multicolumn{1}{|c|}{273}\\

\hline
\end{tabular}
\end{table}

The results obtained for the \gls{DOM} electronics boards are presented in Table~\ref{tab:tabfit}. To fully quantify the reliability of the boards, it is necessary to evaluate each subsystem included in order to exclude, from the total \gls{FIT}, those subsystems that are not critical or do not affect the overall performance of the detector in case of failure. The evaluation is called the \gls{FMECA}. In the case of the Power Board \gls{FMECA} analysis has shown that the failure of the Nanobeacon and piezo power supplies has no impact in the overall physics performances of the KM3NeT detector, because there is enough redundancy. The results obtained by the FIDES method show that the electronics boards in the \gls{DOM} comply with the quality levels required by the KM3NeT Collaboration~\cite{tdr2013}.

% -------------------------------------------------------------------------------------------
\section{Conclusions}\label{sec:end}
% -------------------------------------------------------------------------------------------

In this paper, the electronics front-end and readout system of the KM3NeT telescopes has been presented. The main electronics boards inside the optical modules - the \acrlong{CLB}, the Power Board, the \gls{PMT} bases and the \acrlong{SCB}s - have been described in detail, including a description of the readout architecture of the front-end electronics. A challenging requirement of the readout system is the 1~ns accuracy of the synchronization of the clocks inside the individual optical modules deployed in a water volume of about one cubic kilometer scale. Additional challenge is the power budget of maximal 7 W, including the \gls{HV} of the 31 3-inch \glspl{PMT}. The full chain of the readout electronics has been successfully qualified in situ during a data taking period from May 2014 to July 2015 at a depth of about 3500 m. The qualification has shown that a sustainable synchronization of 1 ns accuracy between the clocks in the individual optical modules has been achieved. Currently, the first deployed \glspl{DU}, using the first batch of mass-produced \gls{DOM} electronics, have been taking data successfully, thus demonstrating the reliability of the KM3NeT front-end and readout electronics system.

\appendix

\begin{multicols}{2}

\printglossary[type=\acronymtype]

\end{multicols}

\appendix 
\acknowledgments
\sloppy
The authors acknowledge the financial support of the funding agencies:
Agence Nationale de la Recherche (contract ANR-15-CE31-0020),
Centre National de la Recherche Scientifique (CNRS), 
Commission Europ\'eenne (FEDER fund and Marie Curie Program),
Institut Universitaire de France (IUF),
IdEx program and UnivEarthS Labex program at Sorbonne Paris Cit\'e (ANR-10-LABX-0023 and ANR-11-IDEX-0005-02),
Paris \^Ile-de-France Region,
France;
Shota Rustaveli National Science Foundation of Georgia (SRNSFG, FR-18-1268),
Georgia;
Deutsche Forschungsgemeinschaft (DFG),
Germany;
The General Secretariat of Research and Technology (GSRT),
Greece;
Istituto Nazionale di Fisica Nucleare (INFN),
Ministero dell'Istruzione, dell'Universit\`a e della Ricerca (MIUR),
PRIN 2017 program (Grant NAT-NET 2017W4HA7S)
Italy;
Ministry of Higher Education, Scientific Research and Professional Training,
Morocco;
Nederlandse organisatie voor Wetenschappelijk Onderzoek (NWO),
the Netherlands;
The National Science Centre, Poland (2015/18/E/ST2/00758);
National Authority for Scientific Research (ANCS),
Romania;
Plan Estatal de Investigaci\'on (refs.\ FPA2015-65150-C3-1-P, -2-P and -3-P, (MINECO/FEDER)), Severo Ochoa Centre of Excellence program (MINECO), Red Consolider MultiDark, (ref. FPA2017-90566-REDC, MINECO) and Prometeo and Grisol\'ia programs (Generalitat Valenciana),
``la Caixa'' Foundation (ID 100010434) through the fellowship LCF/BQ/IN17/11620019, and the European Union's Horizon 2020 research and innovation programme under the Marie Sk\l{}odowska-Curie grant agreement no. 713673,
Spain.

\bibliographystyle{unsrtnat}

%%%%% Biographies of authors %%%%%

\vspace{2ex}\noindent\textbf{Diego Real} is a PhD. candidate and research engineer at Instituto de F\'isica Corpuscular. He received his BS in Electronics in 1996 and his MS in Control and Electronics in 2000, both from the Polytechnic University of Valencia.  He is the author of several publications on electronics. His current research interests include acquisition and synchronization systems for particle physics. He is, since 2013, the Electronics project leader of the KM3NeT telescope and member of the Technical Advisory Board of the GVD-Baikal telescope.

\vspace{2ex}\noindent\textbf{Vincent van Beveren} is a PhD. candidate and embedded software engineer at National Institute for Subatomic Physics Nikhef. He received his BS in computer science and information technology in 2004 from the Hogeschool Utrecht. He has collaborated on multiple publications on various topics. His areas of expertise is in ultra lower power embedded systems, digital signal processing and Internet of Things (IoT).

\vspace{2ex}\noindent\textbf{David Calvo} is a PhD. candidate and research engineer at Instituto de F\'isica Corpuscular of Valencia. He received his MS in Computing in 2006 from University Jaume I, his MS in Electronics in 2009 from University of Valencia and his MS in electronic systems design in 2012 from the Polytechnic University of Valencia. His research interests are focused on the digital electronics, synchronization and readout acquisition systems. He is the author of several publications on electronics.

\listoffigures
\listoftables
\end{spacing}

\end{document}